\documentclass[letterpaper,journal]{IEEEtran}
\usepackage{amsmath,amsfonts}
\usepackage{algorithmic}
\usepackage{algorithm}
\usepackage{array}
\usepackage[caption=false,font=normalsize,labelfont=sf,textfont=sf]{subfig}
\usepackage{textcomp}
\usepackage{stfloats}
\usepackage{url}
\usepackage{verbatim}
\usepackage{graphicx}
\usepackage{xcolor} 
\usepackage{cite}
\begin{document}

\title{Compute-in-Memory Attention: A Time-Domain Analog Softmax Circuit with RC-Tunable Temperature}

\author{Ankur Singh$^{1}$, Ashish Gautam$^{2}$, Shruti R. Kulkarni$^{2}$ and Guojing Cong$^{2}$
        % <-this % stops a space
\thanks{$^{1}$Department of Electrical and Computer Engineering, University of Wisconsin–Madison, Madison, WI, USA. Email: ankur.singh@wisc.edu.}% <-this % stops a space
\thanks{ORNL
Oak Ridge, Tennessee, USA. Emails: \{gautama, kulkarnisr, congg\}@ornl.gov.}}

% The paper headers
\markboth{}%
{Shell \MakeLowercase{\textit{et al.}}: A Sample Article Using IEEEtran.cls for IEEE Journals}

\IEEEpubid{}
% Remember, if you use this you must call \IEEEpubidadjcol in the second
% column for its text to clear the IEEEpubid mark.

\maketitle

\begin{abstract}
Softmax is a key operation in Transformer attention, but its exponentiation and normalization add significant overhead in compute-in-memory (CIM) accelerators, especially when analog attention scores must first be
converted to the digital domain. This work presents a tunable-temperature analog softmax circuit in GlobalFoundries 22-nm fully depleted silicon-on-insulator (FDSOI) technology that operates directly on CIM-generated score voltages without intermediate analog-to-digital
conversion. Each input score is converted into a time-domain event using a shared falling ramp. The corresponding comparator transition samples an RC-decaying reference to generate an exponential weight, which is
then processed by an in-circuit normalization stage. In contrast to analog softmax circuits that rely on transistor weak-inversion behavior for exponentiation, the proposed architecture controls the softmax response through the ramp slope and RC time constant, enabling
programmable effective temperature. The 128-element architecture is evaluated using transistor-level and post-layout extracted simulations, including multi-level input vectors, capacitance variation and mismatch,
process and temperature variation, monte carlo analysis, and shared-interconnect parasitics. The complete 128-element implementation occupies 9453.42~$\mu\mathrm{m}^{2}$ including the shared
global ramp circuitry, while each replicated softmax element occupies 70.2~$\mu\mathrm{m}^{2}$. The circuit achieves a 242.97-ns evaluation latency at 13.44~mW total power, corresponding to 25.5~pJ per output
element. The simultaneous 128-element evaluation achieves an RMSE of 24.46~mV relative to the ideal softmax response. The extracted circuit
characteristics are further incorporated into a MemTorch-based hardware-aware Transformer model, where the proposed softmax achieves a validation loss within 2.5\% of the ideal-softmax baseline.
\end{abstract}


\section{Introduction}

Transformer networks have become the dominant architecture for large language models (LLMs) and sequence-processing tasks due to their self-attention mechanism \cite{1vaswani2017attention}. In scaled dot-product attention, query, key, and value matrices are used to compute
\begin{equation}
\mathrm{Attention}(\mathbf{Q},\mathbf{K},\mathbf{V}) =
\mathrm{Softmax}\left(\frac{\mathbf{Q}\mathbf{K}^{T}}{\sqrt{d_k}}\right)\mathbf{V},
\label{eq:attention}
\end{equation}
where the softmax function converts each attention-score vector into normalized weights. Unlike conventional classifiers, where softmax is typically applied only at the output layer, transformer models use softmax repeatedly across attention heads and layers. As sequence length and model size increase, softmax becomes a non-negligible operation because it requires exponentiation, maximum search for numerical stability, accumulation, and normalization \cite{2stevens2021softermax}.

Compute-in-memory (CIM) architectures are attractive for transformer acceleration because they reduce data movement by performing matrix--vector or matrix--matrix operations close to, or inside, the memory array \cite{3sebastian2020memory}. This property is well matched to the dot-product operations used to generate attention scores. However, integrating softmax with analog CIM remains challenging. A CIM array can naturally generate analog score voltages or currents, but softmax requires a nonlinear exponential operation followed by vector-wide normalization. Conventional approaches therefore often rely on digital or mixed-signal peripheral blocks, such as lookup-based exponentiation, low-precision arithmetic, reciprocal/division units, or online normalization schemes \cite{2stevens2021softermax}. These techniques reduce digital cost but still require dedicated softmax hardware and may introduce conversion overhead when interfaced with analog CIM outputs.

Recent analog in-memory attention work further illustrates this challenge. In \cite{4leroux2025analog}, the attention computation is implemented with analog gain-cell arrays and charge-to-pulse conversion, avoiding intermediate analog-to-digital converters (ADCs) between attention stages. However, instead of implementing softmax directly, the architecture uses a HardSigmoid nonlinearity, since softmax normalization requires an additional interconnection along the sequence dimension, which is difficult to realize efficiently in analog hardware \cite{4leroux2025analog}. This motivates compact analog softmax circuits that can interface directly with CIM-generated score signals while preserving the normalization behavior required by attention.

Direct analog softmax circuits have been investigated using source-coupled transistor arrays. A representative approach implements softmax with bipolar junction transistor (BJT) or NMOS differential structures, where the normalized branch current follows the softmax form~\cite{5sillman2023analog}. For the NMOS implementation, however, the exponential current--voltage relationship is obtained only when the input transistors operate in subthreshold. In this condition, the drain current is exponentially related to $V_{GS}$, and the normalized current ratio approximates softmax with an effective slope determined by $nU_T$, where $n$ is the subthreshold factor and $U_T$ is the thermal voltage~\cite{5sillman2023analog,6enz1995analytical}. This operating condition also restricts the usable input-voltage range, since all input devices must remain within the weak-inversion region for the exponential relationship to remain valid. As the input voltage increases toward moderate or strong inversion, the current--voltage characteristic deviates from the desired exponential dependence, thereby limiting the dynamic range over which accurate softmax behavior can be maintained. Although common threshold-dependent factors cancel in the ideal matched current ratio, practical operation therefore remains constrained by the narrow weak-inversion input range, temperature dependence, leakage, tail-current accuracy, device mismatch, and the need to keep all input devices in the intended operating region.

This work presents a programmable-temperature analog softmax circuit for CIM attention in 22-nm CMOS. Instead of relying on weak-inversion MOS exponential behavior, the proposed approach generates the exponential function in the time domain using a shared voltage ramp, comparator-based event detection, and RC-decay sampling. The effective softmax temperature is controlled by the ramp slope and RC time constant, enabling flexible tuning of the score-to-weight mapping within practical circuit constraints.

The key features and contributions of the proposed time-domain analog
softmax architecture are summarized as follows:

\begin{enumerate}

\item The proposed architecture realizes exponential-weight generation
through score-to-time conversion using a shared ramp and RC-decay
sampling, avoiding reliance on the weak-inversion exponential
characteristics conventionally employed in analog softmax circuits.

\item A programmable-temperature softmax operation is achieved by
controlling the ramp slope and RC time constant. The relationship
between these circuit parameters and the effective softmax selectivity
is analytically derived and validated through transistor-level
simulations.

\item The proposed circuit evaluates a 128-element softmax vector in parallel within 242.97~ns at 25.5~pJ per element, achieving an RMSE of 24.46~mV for simultaneous 128-element operation, with robustness validated under capacitance variation, analog nonidealities, process corners, and monte carlo mismatch.

\item Post-layout extraction quantifies the impact of device and
interconnect parasitics on accuracy and performance. The implemented
softmax element occupies 70.2~$\mu$m$^2$, corresponding to
9453.42~$\mu$m$^2$ for the complete 128-element array.

\item The extracted circuit characteristics are incorporated into a MemTorch-based hardware-aware decoder-only Transformer model, achieving a validation loss of 1.7868 versus 1.7435 for ideal softmax ($\approx$2.5\% degradation) and confirming applicability to Transformer
inference.

\end{enumerate}

The remainder of this paper is organized as follows. Section~II
describes the proposed time-domain analog softmax architecture and its
theoretical formulation. Section~III presents circuit-level validation,
including softmax accuracy, programmable-temperature operation, analog
nonideality analysis, process and monte-carlo characterization, and
post-layout extracted results. Section~IV presents the MemTorch-based
Transformer evaluation, compares the proposed architecture with prior
softmax hardware, and discusses future implementation and
scalability. Finally, Section~V concludes the paper.

\section{Proposed Time-Domain Analog Softmax Circuit}

\subsection{Temperature-Programmable Softmax}

Softmax is a fundamental operation in attention mechanisms that converts raw similarity scores into a probability distribution. It enables the model to emphasize more relevant tokens while suppressing less important ones. Softmax transforms an attention-score vector $\mathbf{s}=[s_1,s_2,\ldots,s_N]$ into normalized attention weights as
\begin{equation}
p_i=\frac{\exp\left(s_i/T\right)}
{\displaystyle\sum_{j=1}^{N}\exp\left(s_j/T\right)},
\qquad i=1,\ldots,N,
\label{eq:softmax}
\end{equation}
where $p_i$ denotes the attention weight corresponding to score $s_i$, and $T$ is the softmax temperature. The normalization ensures that $\sum_{i=1}^{N}p_i=1$, allowing the resulting weights to represent the relative importance of tokens in the attention operation. The temperature controls the selectivity of the attention distribution. A lower $T$ increases the relative weight of the largest scores and produces a sharper distribution, whereas a higher $T$ produces smoother and more uniform attention weights.

The softmax characteristic can equivalently be expressed using the inverse-temperature coefficient, or softmax gain, $\gamma$, defined as
\begin{equation}
\gamma=\frac{1}{T}.
\label{eq:softmax_gain}
\end{equation}
Accordingly, \eqref{eq:softmax} can be rewritten as
\begin{equation}
p_i=
\frac{\exp\left(\gamma s_i\right)}
{\displaystyle\sum_{j=1}^{N}\exp\left(\gamma s_j\right)}.
\label{eq:softmax_gamma}
\end{equation}
A larger $\gamma$ corresponds to a lower effective temperature and produces a sharper probability distribution, whereas a smaller $\gamma$ produces a smoother distribution.

In CIM-based Transformer accelerators, the score range can vary with attention head, layer, array gain, and analog nonidealities. Therefore, a programmable softmax characteristic enables the score-to-weight mapping to be adjusted without modifying the preceding CIM dot-product operation. In the proposed circuit, this programmability is achieved through the slope of a shared ramp and the time constant of an RC-decay waveform. As derived in the following subsection, these circuit parameters directly determine the effective softmax temperature and the corresponding inverse-temperature coefficient $\gamma$.

\begin{figure*}[!t]
    \centering
    \includegraphics[width=0.97\textwidth]{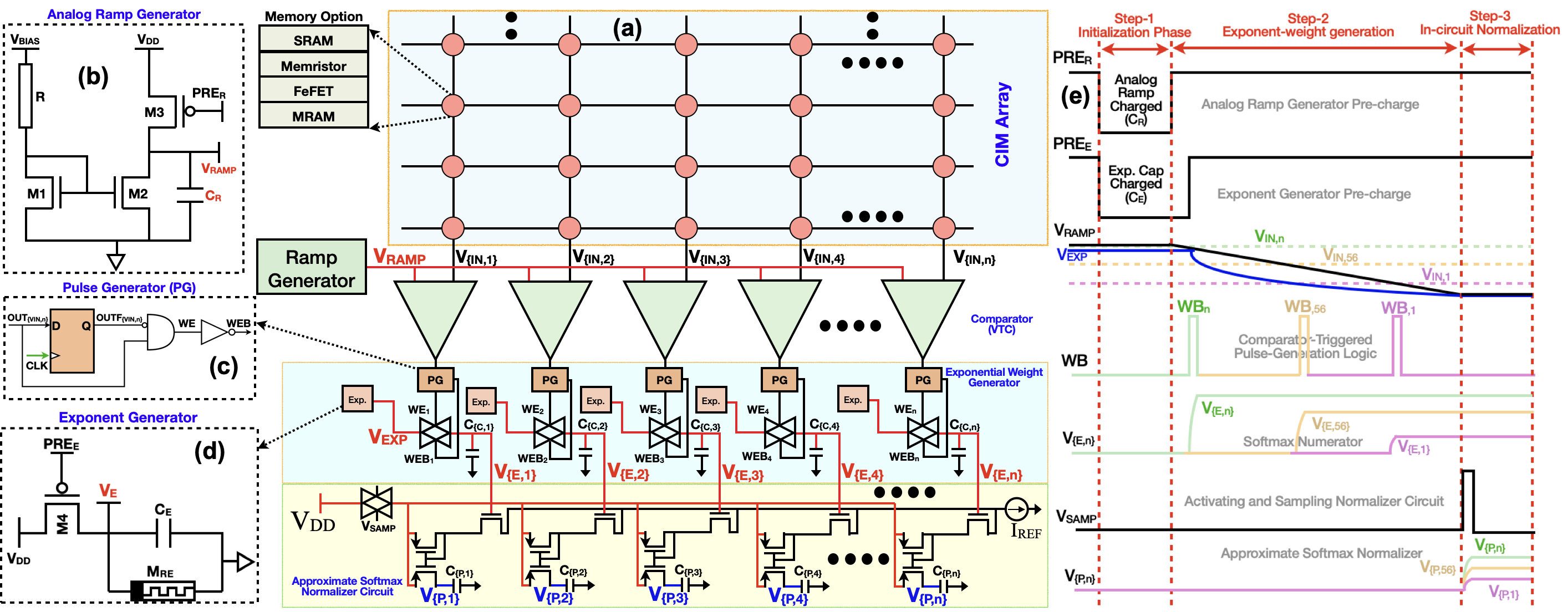}
    \caption{Proposed time-domain analog softmax architecture: (a) CIM array generating the input score voltages $V_{\mathrm{IN},n}$; (b) shared ramp generator and comparator array for score-to-time conversion; (c) comparator-triggered pulse generation and RC-decay sampling to store exponential weights $V_{E,n}$; (d) approximate in-circuit normalizer generating $V_{P,n}$; and (e) timing sequence of the three operating phases.}
    \label{fig:proposed_architecture}
\end{figure*}

\begin{figure}[!t]
    \centering
    \includegraphics[width=0.38\textwidth]{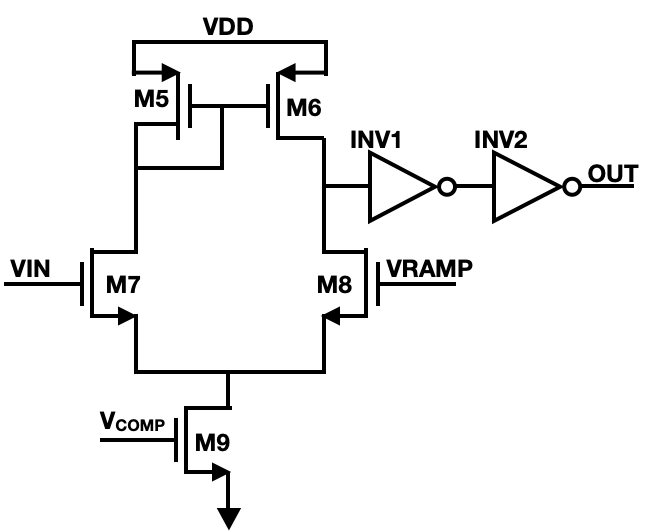}
    \caption{Schematic diagram of the static comparator.}
    \label{fig:comparator}
\end{figure}

\subsection{Time-Domain Analog Softmax Architecture}

Fig.~\ref{fig:proposed_architecture} illustrates the proposed time-domain analog softmax architecture, which comprises a shared ramp generator, a comparator {\cite{baker2019cmos}} array for score-to-time conversion, RC-based exponential-weight generators, local pulse-generation and sampling circuits, and an approximate in-circuit normalization block. The CIM-generated score voltages $V_{\mathrm{IN},1},V_{\mathrm{IN},2},\ldots,V_{\mathrm{IN},N}$ are applied to comparator cells sharing a common falling ramp voltage, $V_{\mathrm{RAMP}}(t)$. Each comparator switches when the ramp reaches its corresponding input voltage.

The resulting comparator transition is processed by the local pulse-generation circuit shown in Fig.~\ref{fig:proposed_architecture}(c) to produce a short complementary sampling pulse, $\mathrm{WE}_n/\mathrm{WEB}_n$, which enables the associated transmission gate. The exponentially decaying reference voltage, $V_{\mathrm{EXP}}(t)$, is then sampled onto the exponent-storage capacitor to generate $V_{E,n}$ for the $n$th input. These stored voltages represent exponential weights and are subsequently applied to the normalization block to generate the output voltages $V_{P,1},V_{P,2},\ldots,V_{P,N}$. The voltage $V_{P,n}$ represents the normalized attention weight corresponding to $V_{\mathrm{IN},n}$.

The circuit operation, illustrated by the timing sequence in Fig.~\ref{fig:proposed_architecture}(e), is divided into three phases: 1) initialization, 2) exponential-weight generation, and 3) approximate in-circuit normalization.

\textit{1) Initialization phase:}
At the beginning of each softmax operation, the active-low PMOS precharge signals $\mathrm{PRE_R}$ and $\mathrm{PRE_E}$ are asserted. The ramp capacitor $C_R$ and exponential-generator capacitor $C_E$ are consequently charged to approximately $V_{\mathrm{DD}}$. The memristive element \cite{16li2026high, 17singh2021current} in the exponential generator is programmed before inference to a high-resistance state, denoted by $R_{\mathrm{HRS}}$, such that the required RC time constant is obtained. This programming operation is performed only when the desired softmax operating condition is configured and is not repeated for every input vector. Simultaneously, the exponent-storage capacitors $C_{C,n}$ and probability-output capacitors $C_{P,n}$ are reset to 0~V.

After the initialization interval, $\mathrm{PRE_R}$ and $\mathrm{PRE_E}$ are deasserted, turning off the PMOS precharge devices. The ramp capacitor is discharged through a bias-controlled current path. Assuming an approximately constant discharge current $I_R(V_{\mathrm{BIAS}})$, the falling ramp is expressed as
\begin{equation}
V_{\mathrm{RAMP}}(t)=V_{\mathrm{DD}}-S_R(t-t_0),
\quad
S_R=\frac{I_R(V_{\mathrm{BIAS}})}{C_R},
\label{eq:ramp_voltage}
\end{equation}
where $t_0$ denotes the beginning of the evaluation interval and $S_R$ is the ramp slope. Therefore, $V_{\mathrm{BIAS}}$ provides electrical control of the ramp slope.

At the same time, $C_E$ discharges through the programmed memristive resistance. Assuming that the resistance remains approximately constant over the selected operating range, the exponentially decaying voltage is expressed as
\begin{equation}
V_{\mathrm{EXP}}(t)=V_{\mathrm{DD}}
\exp\left(-\frac{t-t_0}{\tau_E}\right),
\quad
\tau_E=R_{\mathrm{HRS}}C_E,
\label{eq:exp_reference}
\end{equation}
where $\tau_E$ denotes the exponential-generator time constant.

\textit{2) Exponential-weight generation phase:}
The input voltages $V_{\mathrm{IN},1},V_{\mathrm{IN},2},\ldots,V_{\mathrm{IN},N}$ correspond to the analog attention-score outputs of a preceding CIM dot-product operation \cite{7chakraborty2026gem3d,8gi2025memristor}. In this work, $V_{\mathrm{IN},n}$ denotes the analog score voltage from the preceding CIM stage, whose detailed circuit implementation is outside the scope of this work.

Each score voltage $V_{\mathrm{IN},n}$ is compared with the common falling ramp. The $n$th comparator switches when
\begin{equation}
V_{\mathrm{RAMP}}(t_n)=V_{\mathrm{IN},n}.
\end{equation}
Substituting \eqref{eq:ramp_voltage}, the corresponding switching time is
\begin{equation}
t_n-t_0=
\frac{V_{\mathrm{DD}}-V_{\mathrm{IN},n}}{S_R}.
\label{eq:switching_time}
\end{equation}
Hence, a larger input score produces an earlier comparator transition.

The comparator output is processed by the local pulse-generation circuit to produce a narrow complementary sampling pulse, $\mathrm{WE}_n/\mathrm{WEB}_n$. The pulse turns on the corresponding transmission gate and samples $V_{\mathrm{EXP}}(t)$ onto $C_{C,n}$. For a sampling pulse width $T_W$ satisfying
\begin{equation}
R_{\mathrm{TG}}C_{C,n}\ll T_W\ll\tau_E,
\label{eq:sampling_condition}
\end{equation}
where $R_{\mathrm{TG}}$ denotes the on-resistance of the transmission gate, the sampled voltage can be approximated as
\begin{equation}
V_{E,n}\approx V_{\mathrm{EXP}}(t_n).
\end{equation}
Substituting \eqref{eq:switching_time} into \eqref{eq:exp_reference} gives
\begin{equation}
V_{E,n}=V_{\mathrm{DD}}\exp\left(
-\frac{V_{\mathrm{DD}}-V_{\mathrm{IN},n}}
{S_R\tau_E}
\right).
\label{eq:sampled_weight}
\end{equation}

Equation \eqref{eq:sampled_weight} can be rewritten as
\begin{equation}
V_{E,n}=A_E
\exp\left(
\frac{V_{\mathrm{IN},n}}{T_{\mathrm{eff}}}
\right),
\label{eq:exp_weight}
\end{equation}
where
\begin{equation}
A_E=
V_{\mathrm{DD}}
\exp\left(
-\frac{V_{\mathrm{DD}}}{S_R\tau_E}
\right)
\end{equation}
is common to all branches, and
\begin{equation}
T_{\mathrm{eff}}=S_R\tau_E.
\label{eq:effective_temperature}
\end{equation}

Since $A_E$ is common to all branches, the sampled voltage is proportional to the exponential softmax numerator:
\begin{equation}
V_{E,n}\propto
\exp\left(
\frac{V_{\mathrm{IN},n}}{T_{\mathrm{eff}}}
\right).
\end{equation}

Using $\tau_E=R_{\mathrm{HRS}}C_E$, the effective temperature can be written explicitly in terms of circuit parameters as
\begin{equation}
T_{\mathrm{eff}}
=
S_RR_{\mathrm{HRS}}C_E.
\label{eq:teff_expanded}
\end{equation}
Furthermore, since $S_R=I_R(V_{\mathrm{BIAS}})/C_R$,
\begin{equation}
T_{\mathrm{eff}}
=
\frac{I_R(V_{\mathrm{BIAS}})}
{C_R}
R_{\mathrm{HRS}}C_E.
\label{eq:teff_circuit}
\end{equation}

Therefore, the exponential mapping can be programmed through either the ramp slope $S_R$ or the RC time constant $\tau_E$. The ramp bias provides dynamic electrical control of the score-to-time conversion, while the programmable resistance provides an additional means of selecting the RC operating point. A smaller $T_{\mathrm{eff}}$ produces a sharper exponential response, whereas a larger $T_{\mathrm{eff}}$ produces a smoother response.

\textit{3) Approximate in-circuit normalization phase:}
After the exponent-storage capacitors retain their sampled voltages,
$V_{E,1},V_{E,2},\ldots,V_{E,N}$ are applied to the gates of
matched NMOS transistors. The source terminals of all $N$ NMOS
transistors are connected to a common node terminated by the reference
current sink $I_{\mathrm{REF}}$. For analysis, the voltage developed at
this common source node is denoted by $V_S$. Each NMOS drain is
connected to the input branch of a PMOS current mirror, while the
corresponding mirror-output branch charges the probability capacitor
$C_{P,n}$. The PMOS source rail is connected to $V_{\mathrm{DD}}$
through the $V_{\mathrm{SAMP}}$-controlled switch. After all
exponential weights have been stored, $V_{\mathrm{SAMP}}$ is asserted
for the normalization interval, enabling the current-mirror network.

For the $n$th NMOS transistor,
\begin{equation}
V_{GS,n}=V_{E,n}-V_S,
\qquad
V_{DS,n}=V_{D,n}-V_S,
\label{eq:norm_terminal}
\end{equation}
where $V_{D,n}$ denotes the corresponding NMOS drain voltage. The
saturation condition is
\begin{equation}
V_{DS,n}\geq V_{GS,n}-V_{\mathrm{TH}},
\end{equation}
or equivalently,
\begin{equation}
V_{D,n}\geq V_{E,n}-V_{\mathrm{TH}}.
\label{eq:norm_sat_condition}
\end{equation}

Omitting channel-length modulation in the first-order analytical
model, the branch current is
\begin{equation}
I_n
\approx
\frac{\beta}{2}
\left(
V_{E,n}-V_S-V_{\mathrm{TH}}
\right)^2,
\label{eq:norm_branch_current}
\end{equation}
where
\begin{equation}
\beta=\mu_n C_{\mathrm{ox}}\frac{W}{L}.
\end{equation}

Since all NMOS source terminals share the same node, Kirchhoff's current
law gives
\begin{equation}
I_{\mathrm{REF}}
=
\sum_{k=1}^{N} I_k.
\label{eq:iref_sum}
\end{equation}
Defining the common offset
\begin{equation}
V_0=V_S+V_{\mathrm{TH}},
\end{equation}
the normalized branch-current ratio can be written as
\begin{equation}
\frac{I_n}{I_{\mathrm{REF}}}
\approx
\frac{
\left(V_{E,n}-V_0\right)^2
}{
\displaystyle
\sum_{k=1}^{N}
\left(V_{E,k}-V_0\right)^2
}.
\label{eq:norm_current_ratio}
\end{equation}

From the exponential-weight generation phase,
\begin{equation}
V_{E,n}
=
A_E
\exp\left(
\frac{V_{\mathrm{IN},n}}{T_{\mathrm{eff}}}
\right).
\label{eq:norm_exp_weight}
\end{equation}
Therefore,
\begin{equation}
\frac{I_n}{I_{\mathrm{REF}}}
\approx
\frac{
\left[
A_E
\exp\left(
\dfrac{V_{\mathrm{IN},n}}{T_{\mathrm{eff}}}
\right)-V_0
\right]^2
}{
\displaystyle
\sum_{k=1}^{N}
\left[
A_E
\exp\left(
\dfrac{V_{\mathrm{IN},k}}{T_{\mathrm{eff}}}
\right)-V_0
\right]^2
}.
\label{eq:norm_exact_approx}
\end{equation}

For a first-order closed-form relation between the circuit parameters
and the softmax temperature, the common offset $V_0$ is neglected in
the dominant input-dependent term. The normalized current ratio then
reduces to
\begin{equation}
\frac{I_n}{I_{\mathrm{REF}}}
\approx
\frac{
\exp\left(
\dfrac{2V_{\mathrm{IN},n}}{T_{\mathrm{eff}}}
\right)
}{
\displaystyle
\sum_{k=1}^{N}
\exp\left(
\dfrac{2V_{\mathrm{IN},k}}{T_{\mathrm{eff}}}
\right)
}.
\label{eq:norm_softmax_ratio}
\end{equation}

Therefore, the square-law normalization stage introduces an additional
factor of two in the exponential coefficient. The effective softmax
temperature at the normalized output is
\begin{equation}
T_{\mathrm{sat}}
=
\frac{T_{\mathrm{eff}}}{2}
=
\frac{S_R\tau_E}{2}.
\label{eq:norm_sat_temperature}
\end{equation}
The corresponding theoretical inverse-temperature coefficient is
\begin{equation}
\gamma_{\mathrm{th}}
=
\frac{1}{T_{\mathrm{sat}}}
=
\frac{2}{S_R\tau_E}.
\label{eq:gamma_theoretical}
\end{equation}

Using $\tau_E=R_{\mathrm{HRS}}C_E$,
\begin{equation}
\gamma_{\mathrm{th}}
=
\frac{2}
{S_RR_{\mathrm{HRS}}C_E},
\label{eq:gamma_rc}
\end{equation}
and since $S_R=I_R(V_{\mathrm{BIAS}})/C_R$,
\begin{equation}
\gamma_{\mathrm{th}}
=
\frac{2C_R}
{I_R(V_{\mathrm{BIAS}})
R_{\mathrm{HRS}}C_E}.
\label{eq:gamma_circuit}
\end{equation}

Equations~\eqref{eq:gamma_theoretical}--\eqref{eq:gamma_circuit}
directly relate the softmax selectivity to the ramp and RC operating
conditions. Increasing $S_R$ increases the effective temperature and
reduces $\gamma_{\mathrm{th}}$, whereas increasing the RC time constant
also reduces $\gamma_{\mathrm{th}}$. The finite common-source voltage,
threshold voltage, channel-length modulation, and current-mirror
nonidealities are retained in the transistor-level simulations and
therefore contribute to the difference between the theoretical and
simulated softmax responses.

During the $V_{\mathrm{SAMP}}$ interval, the PMOS current mirrors
transfer the NMOS branch currents to the corresponding output
capacitors. For matched 1:1 current mirrors,
\begin{equation}
I_{P,n}\approx I_n,
\end{equation}
where $I_{P,n}$ denotes the mirrored current charging $C_{P,n}$. The
output capacitors are initially reset such that $V_{P,n}\approx0$, and
during the normalization interval,
\begin{equation}
C_{P,n}
\frac{\mathrm{d}V_{P,n}}{\mathrm{d}t}
=
I_{P,n}.
\end{equation}

Integrating over the $V_{\mathrm{SAMP}}$ pulse width
$T_{\mathrm{SAMP}}$ gives
\begin{equation}
V_{P,n}
=
\frac{1}{C_{P,n}}
\int_{0}^{T_{\mathrm{SAMP}}}
I_{P,n}(t)\,\mathrm{d}t.
\label{eq:norm_output_integration}
\end{equation}

For equal output capacitors and an approximately constant current ratio
during the short normalization interval,
\begin{equation}
V_{P,n}
\approx
V_{\mathrm{FS}}
\frac{
\exp\left(
\dfrac{V_{\mathrm{IN},n}}{T_{\mathrm{sat}}}
\right)
}{
\displaystyle
\sum_{k=1}^{N}
\exp\left(
\dfrac{V_{\mathrm{IN},k}}{T_{\mathrm{sat}}}
\right)
},
\qquad
V_{\mathrm{FS}}
=
\frac{I_{\mathrm{REF}}T_{\mathrm{SAMP}}}{C_P}.
\label{eq:norm_output_softmax}
\end{equation}

Using $\gamma_{\mathrm{th}}=1/T_{\mathrm{sat}}$, the output relation can
equivalently be expressed as
\begin{equation}
V_{P,n}
\approx
V_{\mathrm{FS}}
\frac{
\exp\left(
\gamma_{\mathrm{th}}V_{\mathrm{IN},n}
\right)
}{
\displaystyle
\sum_{k=1}^{N}
\exp\left(
\gamma_{\mathrm{th}}V_{\mathrm{IN},k}
\right)
}.
\label{eq:output_gamma}
\end{equation}

Thus, apart from the common output-scaling factor $V_{\mathrm{FS}}$,
the proposed circuit realizes an approximate vector-wise normalized
exponential response whose first-order inverse-temperature coefficient
is determined by the ramp slope and RC time constant. The effects of
the neglected circuit nonidealities are quantified through the
transistor-level and post-layout simulations in Section~III.

\section{Results and Discussion}

\subsection{Softmax Accuracy and Programmable Temperature}

The viability of the proposed softmax architecture is evaluated using
the cadence virtuoso analog design environment in the
GlobalFoundries 22-nm FDSOI technology~\cite{18carter201622nm}. All active
and passive circuit components, including transistors, capacitors, and
resistors, are implemented using the corresponding technology models.
The nominal simulations use $V_{\mathrm{DD}}=1.1$~V, a ramp capacitance
of $C_R=200$~fF, $V_{\mathrm{BIAS}}=0.34$~V, an
exponential-generator capacitance of $C_E=25$~fF, an exponent-storage
capacitance of $C_C=2$~fF, a probability-output capacitance of
$C_P=2$~fF, and a flip-flop clock period of 200~ps. All capacitances
are implemented using metal--oxide--metal (MOM) capacitors available in
the technology. The transistor dimensions of the comparator circuit shown in Fig.~\ref{fig:comparator} are summarized in Table~\ref{tab:comparator_dimensions}. The programmable resistive element
is configured to a nominal high-resistance state of
$R_{\mathrm{HRS}}=1.5$\,M$\Omega$. From the nominal transient response,
the ramp slope $S_R$ is extracted directly from
$V_{\mathrm{RAMP}}(t)$, while the exponential-decay time constant is
determined from $\tau_E=R_{\mathrm{HRS}}C_E$. Using the
relationship derived in Section~II, the theoretical inverse-temperature
coefficient is
\begin{equation}
\gamma_{\mathrm{th}}=
\frac{2}{S_R\tau_E}.
\label{eq:gamma_sim_validation}
\end{equation}

\begin{table}[!t]
\centering
\caption{Transistor Dimensions of the Comparator Circuit}
\label{tab:comparator_dimensions}
\renewcommand{\arraystretch}{1.15}
\begin{tabular}{lcc}
\hline
\textbf{Device} & \textbf{Width ($W$)} & \textbf{Length ($L$)} \\
\hline
$M_5$--$M_6$ & 4~$\mu$m   & 50~nm  \\
$M_7$--$M_9$ & 300~nm    & 100~nm \\
$INV1,INV2$        & \multicolumn{2}{c}{GF22 standard-cell library} \\
\hline
\end{tabular}
\end{table}

For the nominal operating condition, the circuit parameters used to
determine the theoretical softmax gain are summarized in
Table~\ref{tab:gamma_theory}. The extracted ramp slope of
$S_R=4.501$~mV/ns and RC time constant of $\tau_E=37.5$~ns yield a
theoretical inverse-temperature coefficient of
$\gamma_{\mathrm{th}}=11.85$~V$^{-1}$. Independently, fitting the
transistor-level softmax response yields
$\gamma_{\mathrm{sim}}=11.36$~V$^{-1}$.

\begin{table}[!t]
\centering
\caption{Nominal Circuit Parameters and Theoretical Softmax Gain}
\label{tab:gamma_theory}
\renewcommand{\arraystretch}{1.15}
\begin{tabular}{lc}
\hline
\textbf{Parameter} & \textbf{Nominal Value} \\
\hline
Extracted ramp slope, $S_R$                  & 4.501~mV/ns \\
Programmed resistance, $R_{\mathrm{HRS}}$    & 1.5~M$\Omega$ \\
Exponential capacitance, $C_E$               & 25~fF \\
RC time constant, $\tau_E$                   & 37.5~ns \\
Effective softmax temperature, $T_{\mathrm{sat}}$ & 84.4~mV \\
Theoretical softmax gain, $\gamma_{\mathrm{th}}$ & 11.85~V$^{-1}$ \\
\hline
\end{tabular}
\end{table}

\begin{figure}[!t]
    \centering
    \includegraphics[width=0.48\textwidth]{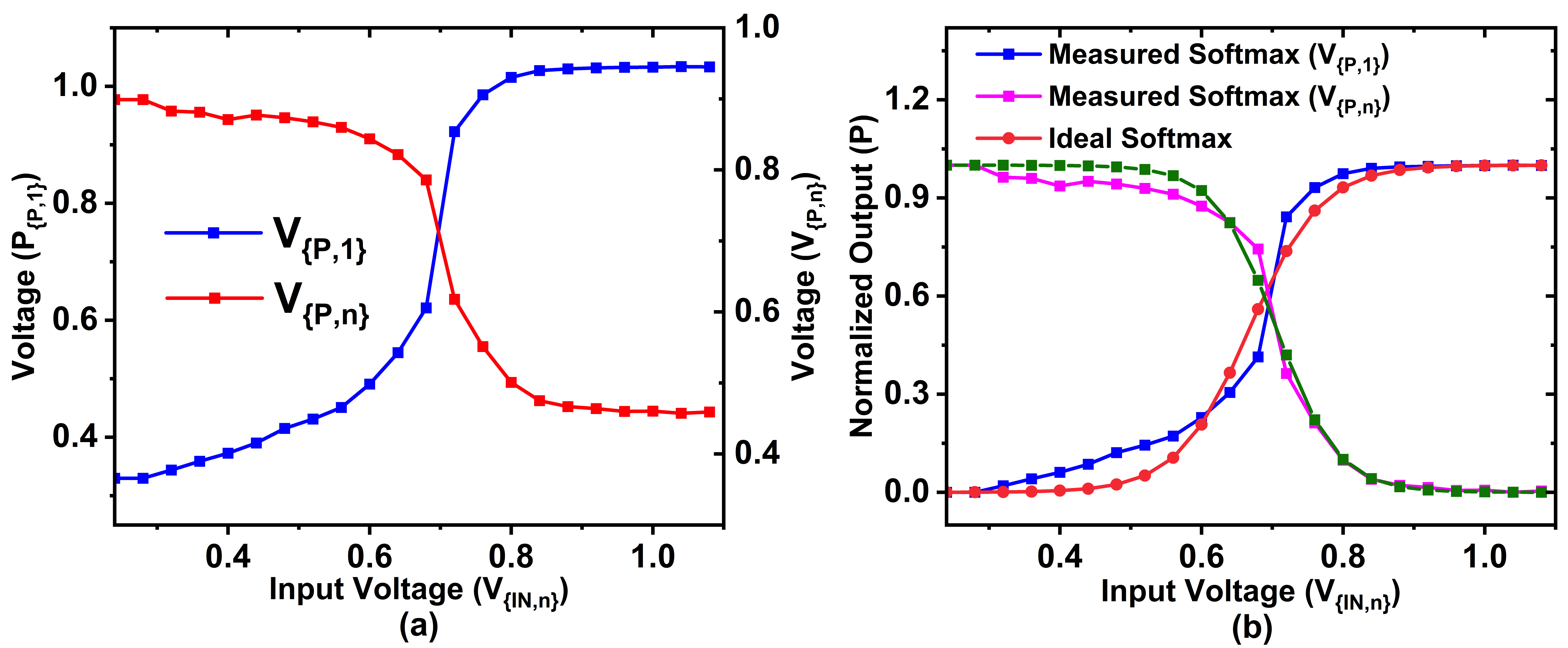}
        \caption{Simulated softmax characteristics: (a) measured transfer, (b) nominal transfer.}
    \label{fig:softmax_transient}
\end{figure}

To first characterize the fundamental transfer behavior of the proposed
normalizer, $V_{\mathrm{IN},1}$ is swept while the remaining 127 input
voltages, $V_{\mathrm{IN},2}$--$V_{\mathrm{IN},128}$, are maintained at
$0.7$\,V. Fig.~\ref{fig:softmax_transient}(a) shows the corresponding simulated
output voltages $V_{P,1}$ and $V_{P,n}$, where $V_{P,n}$ denotes the
output of a representative branch $n\in\{2,\ldots,128\}$ receiving the
fixed $0.7$ V input. As $V_{\mathrm{IN},1}$ increases, $V_{P,1}$
increases while $V_{P,n}$ decreases, and the two responses intersect
near the equal-input condition of $V_{\mathrm{IN},1}=0.7$\,V. This
complementary behavior confirms that the output branches compete through
a common normalization denominator rather than behaving as independent
nonlinear transfer functions. Fig.~\ref{fig:softmax_transient}(b) compares the corresponding normalized transistor-level response with the ideal softmax characteristic using
the nominal circuit-derived inverse-temperature coefficient. The
simulated outputs exhibit the expected monotonic and complementary
softmax behavior over the input-voltage range and achieve an RMSE of
approximately $4.3\%$ with respect to the ideal response. This
controlled single-input experiment provides a direct characterization
of the nominal softmax transfer and common-denominator normalization.

\begin{figure}[!t]
    \centering
    \includegraphics[width=0.48\textwidth]{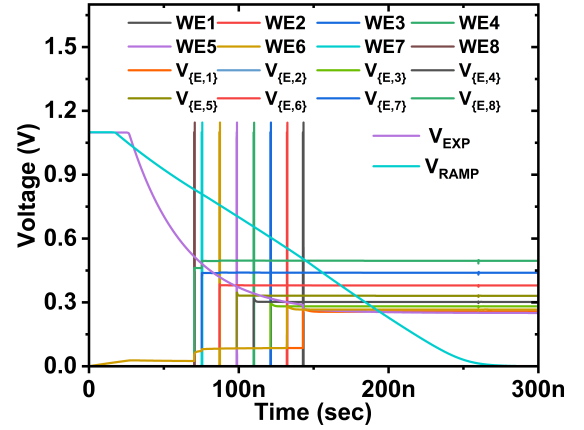}
    \caption{Transient operation for the eight representative input
    levels used in the 128-element vector evaluation, showing the shared
    $V_{\mathrm{RAMP}}$ and $V_{\mathrm{EXP}}$ waveforms, the
    corresponding write-enable signals $WE_1$--$WE_8$, and the sampled
    voltages $V_{E,1}$--$V_{E,8}$.}
    \label{fig:vector_transient}
\end{figure}

To evaluate the vector-level softmax accuracy under different input
conditions, the proposed circuit was simulated with a 128-element
interleaved multi-level input vector instead of the controlled
single-input sweep described above. Eight voltage levels,
0.50, 0.55, 0.60, 0.65, 0.70, 0.75, 0.80, and 0.82~V, were used to
activate all 128 input channels simultaneously. Each voltage level was
applied to 16 channels and interleaved across the 128-channel array.
Specifically, channels $8k$ received 0.50~V, channels $8k+1$ received
0.55~V, and so forth, up to channels $8k+7$ receiving 0.82~V, for
$k=0,\ldots,15$. Fig.~\ref{fig:vector_transient} illustrates the corresponding transient
operation for the eight representative input levels. The shared
$V_{\mathrm{RAMP}}$ waveform generates different crossing times for
the eight inputs, resulting in the corresponding write-enable signals
$WE_1$--$WE_8$. These signals sample the common exponential waveform
$V_{\mathrm{EXP}}$ at different instants and store the resulting
voltages at $V_{E,1}$--$V_{E,8}$. Earlier sampling therefore captures
a larger $V_{\mathrm{EXP}}$ value, whereas later sampling produces a
smaller stored voltage. Since the same eight-level input sequence is
repeated over the complete array, these transient waveforms represent
the eight operating conditions simultaneously distributed across the
128 softmax channels.

\begin{figure}[!t]
    \centering
    \includegraphics[width=0.48\textwidth]{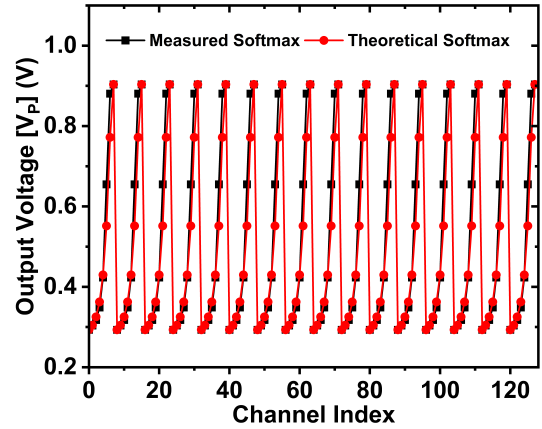}
    \caption{Vector-level validation of the proposed softmax circuit: applied 128-element multi-level input-score vector.}
    \label{fig:vector128}
\end{figure}

The resulting transistor-level outputs of all 128 channels are shown
in Fig.~\ref{fig:vector128} and compared with the corresponding ideal
softmax values calculated using the circuit-derived inverse-temperature
coefficient. The proposed circuit preserves the relative ordering and
normalized distribution of the multi-level input vector, with an
overall RMSE of approximately 24.46~mV over the 128 elements with
respect to the ideal softmax response. These results demonstrate
vector-wise normalization under simultaneous multi-element excitation
and validate the scalability of the proposed architecture to
128-element softmax operation.

\subsection{Analog Non-idealities}

Analog nonidealities can have a significant impact \cite{20ronchi2025design} on the critical
storage nodes of the proposed circuit, particularly at $V_{E,n}$ and
$V_{P,n}$, as well as at the associated transmission-gate interfaces.
Although the nominal storage and probability-output capacitances are
$C_C=C_P=2$~fF, the effective capacitance at these nodes also includes
contributions from interconnect, transmission-gate parasitics,
connected-device capacitances, and capacitive coupling {\cite{21wu20264}}. Accordingly,
the effective capacitance can be expressed as
\begin{equation}
C_{\mathrm{eff}}
=
C_{\mathrm{des}}
+
C_{\mathrm{wire}}
+
C_{\mathrm{TG}}
+
C_{\mathrm{device}}
+
C_{\mathrm{coupling}}
\end{equation}
where $C_{\mathrm{des}}$ denotes the intended capacitance. Since these
parasitic components can become comparable to the nominal capacitance at
the femtofarad scale, they can alter the sampled exponential voltage and
the subsequent normalization process, thereby increasing the deviation
from the theoretical softmax response.

To evaluate this sensitivity, $C_C$ and $C_P$ were independently varied
over a $\pm25\%$ range while applying the same 128-element input vector
used for the nominal characterization. The resulting softmax outputs
were compared with the theoretical reference and the corresponding RMSE
presented. As expected, larger deviations from the nominal
capacitance increase the output error. For instance, increasing $C_P$ to
2.5~fF results in an RMSE of approximately 110~mV with respect to the
nominal theoretical softmax response. However, this capacitance-induced
shift can be compensated by recalibrating the ramp slope, since the
ramp slope directly determines the theoretical inverse-temperature
coefficient $\gamma_{\mathrm{th}}$ and hence the effective softmax
temperature. Therefore, the programmable ramp provides a means of
recovering the desired softmax characteristic in the presence of
systematic capacitance variation. 

\begin{figure}[!t]
    \centering
    \includegraphics[width=0.42\textwidth]{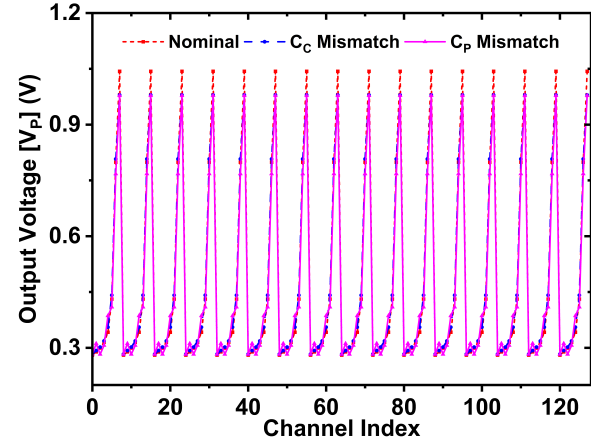}
    \caption{Effect of element-to-element capacitance mismatch on the 128-channel softmax response. Eight capacitance values are distributed across groups of 16 elements for separate $C_C$ and $C_P$ mismatch evaluations and compared with the nominal $2$-fF case.}
    \label{fig:cap_mismatch}
\end{figure}

In addition to the uniform capacitance sweep, element-to-element capacitance mismatch was evaluated at the array level. The 128 softmax elements were divided into eight capacitor groups, with 16 elements assigned to each capacitance value. Starting from the nominal $2$ fF capacitance, eight values spanning approximately
$1.8$--$2.37$~fF were assigned across the array. Two separate simulations were performed: in the first, $C_C$ was varied according to the assigned group values while $C_P$ was maintained at its nominal value, whereas in the second, the same mismatch profile was applied to $C_P$ while $C_C$ was kept nominal. The same 128-element interleaved
input vector was applied in all cases. Fig.~\ref{fig:cap_mismatch}
compares the resulting $C_C$ and $C_P$ mismatched outputs with the
nominal case in which $C_C=C_P=2$~fF. The grouped mismatch produces an
RMSE of approximately $23.6$~mV for $C_C$ variation and $32.7$~mV for
$C_P$ variation relative to the nominal 128-channel response,
demonstrating the propagation of local capacitance variation to the
final normalized softmax output.

\begin{figure}[!t]
    \centering
    \includegraphics[width=0.48\textwidth]{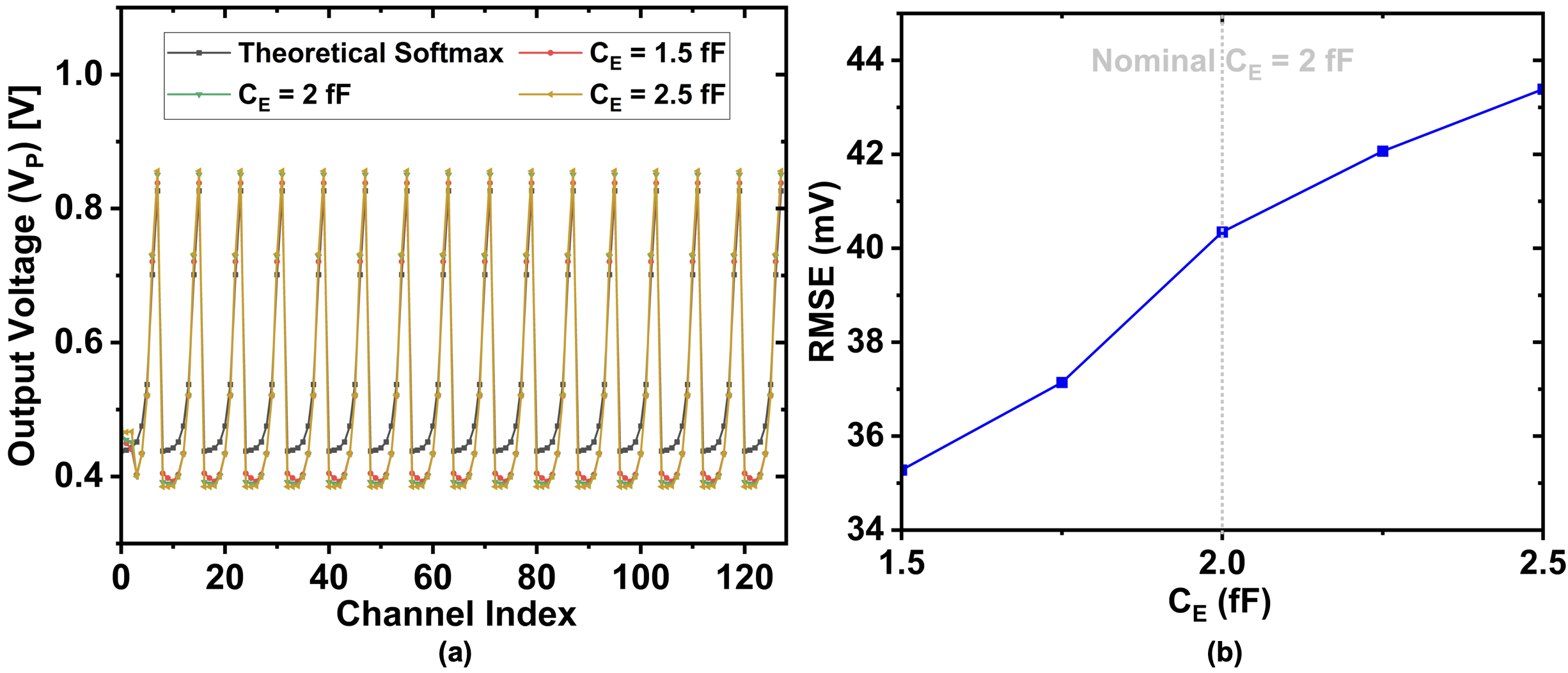}
    \caption{Sensitivity of the proposed softmax circuit of $C_{C,n}$
    (a) 128-element output response for different capacitance values and
    (b) corresponding RMSE with respect to the theoretical softmax reference.}
    \label{fig:softmax_cap_CE}
\end{figure}

\begin{figure}[!t]
    \centering
    \includegraphics[width=0.48\textwidth]{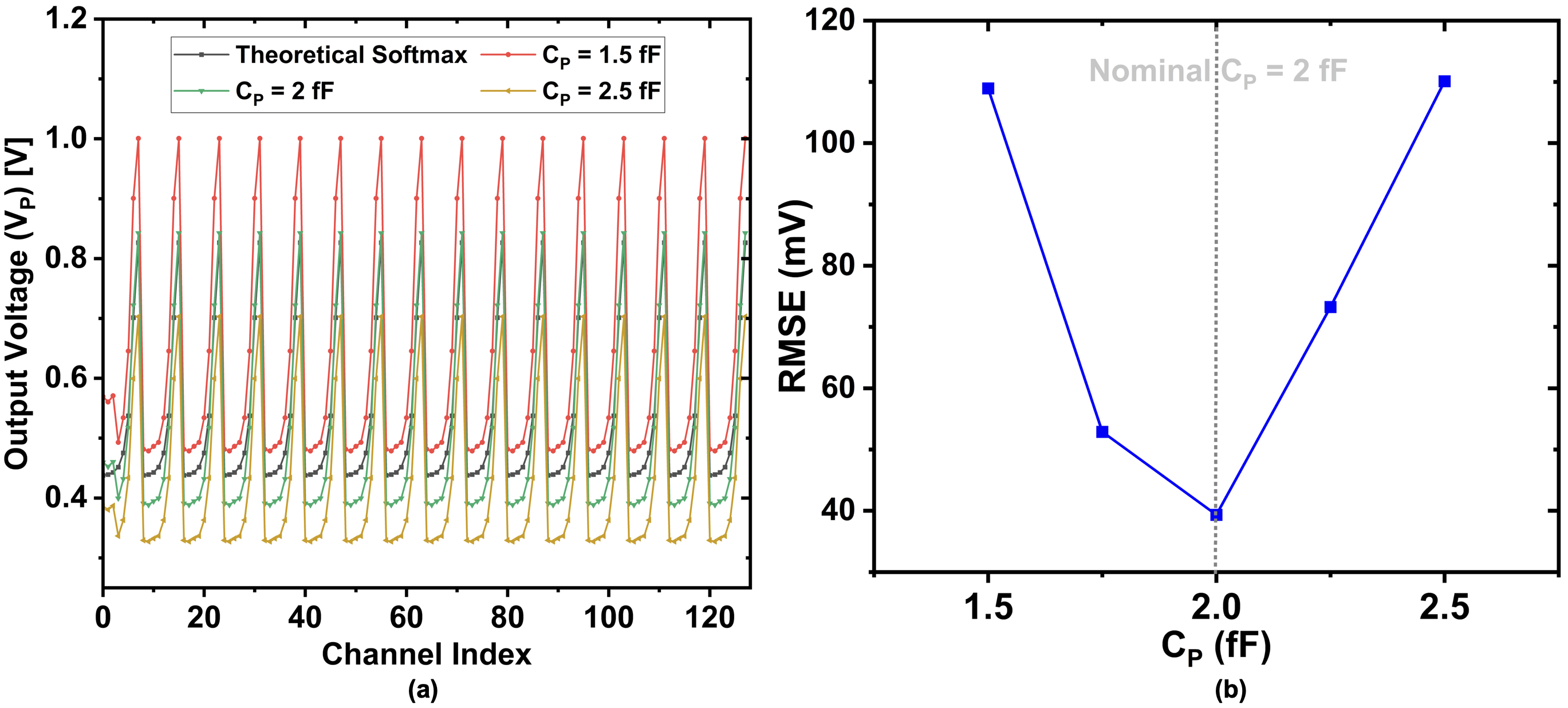}
    \caption{Sensitivity of the proposed softmax circuit of $C_{P,n}$
    (a) 128-element output response for different capacitance values and
    (b) corresponding RMSE with respect to the theoretical softmax reference.}
    \label{fig:softmax_cap_CP}
\end{figure}

\begin{figure}[!t]
    \centering
    \includegraphics[width=0.42\textwidth]{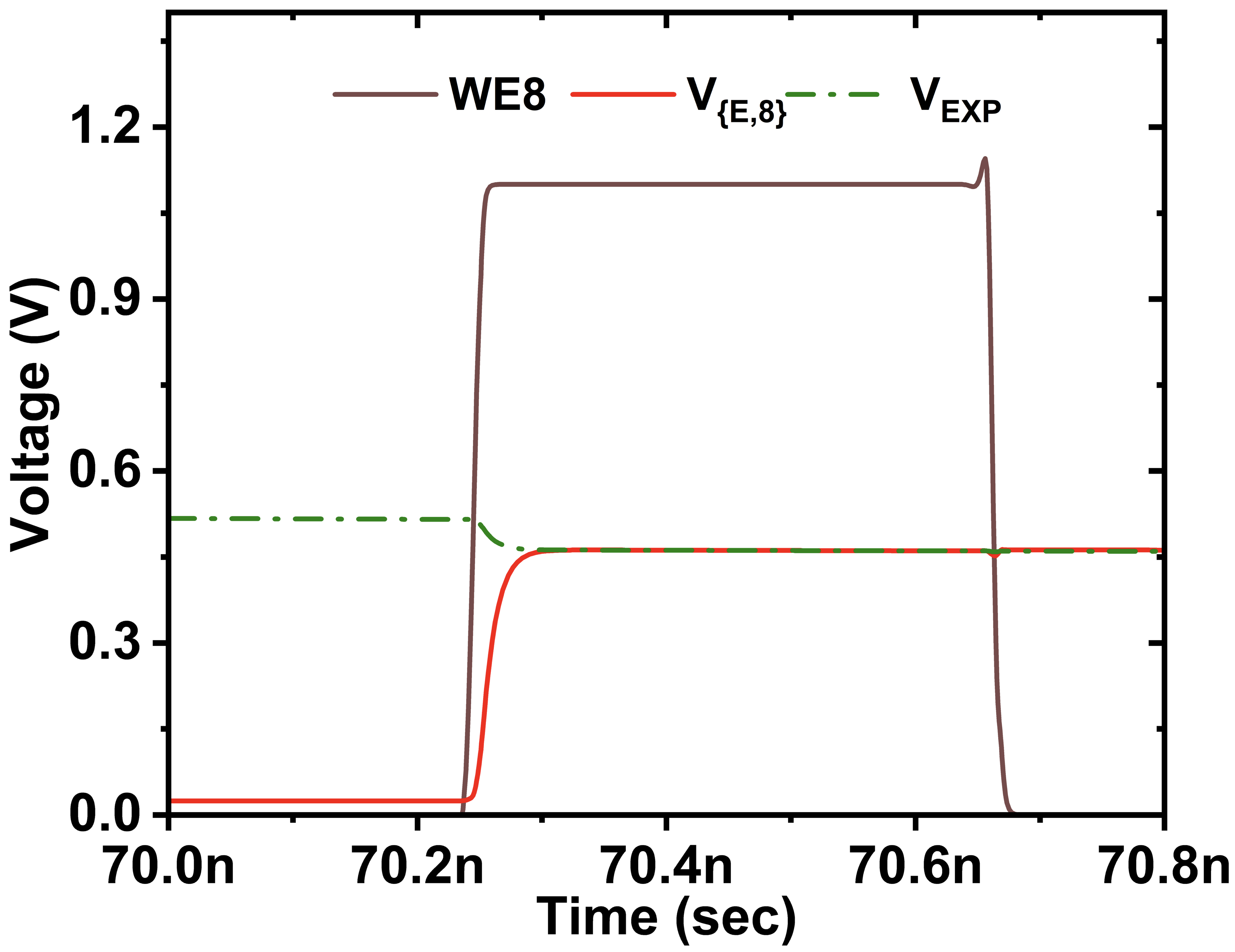}
    \caption{Magnified transient response around the turn-off transition of $WE_8$, showing $V_{\mathrm{EXP}}$ and the corresponding stored voltage $V_{E,8}$ for evaluating transmission-gate charge injection.}
    \label{fig:charge_injection}
\end{figure}

The effect of transmission-gate charge injection was also examined at
the small-capacitance exponential storage node. Fig.~\ref{fig:charge_injection}
shows a magnified transient response around the turn-off transition of
$WE_8$. During the active sampling interval, $V_{E,8}$ follows the
common $V_{\mathrm{EXP}}$ waveform. When $WE_8$ is deasserted, the
transmission gate is disconnected and the sampled voltage is retained
at $V_{E,8}$. A small voltage perturbation is observed at the storage
node during the switching transition due to charge injection. The
resulting sampling error is evaluated from the difference between the
intended $V_{\mathrm{EXP}}$ value at the switching instant and the
subsequently held $V_{E,8}$ voltage.

The globally shared $V_{\mathrm{RAMP}}$ network is also susceptible to
interconnect loading and distributed resistance {\cite{22rack2025high}}. Since the ramp is distributed to all comparator branches, the effective capacitance seen by the ramp generator can be expressed as

\begin{equation}
C_{\mathrm{R,eff}}
=
C_R + C_{\mathrm{RAMP,par}}
\end{equation}

where $C_R$ is the designed ramp capacitance and
$C_{\mathrm{RAMP,par}}$ includes the contributions from the global
interconnect, via connections, and comparator-input loading. The
corresponding effective ramp slope is therefore approximately

\begin{equation}
S_{\mathrm{R,eff}}
\approx
\frac{I_R(V_{\mathrm{BIAS}})}
{C_R+C_{\mathrm{RAMP,par}}}.
\end{equation}

Since the inverse-temperature coefficient is given by
$\gamma_{\mathrm{th}}=2/(S_R\tau_E)$, an increase in the effective ramp
capacitance reduces the ramp slope and consequently modifies the softmax
temperature. This common-mode shift can be compensated by adjusting
$V_{\mathrm{BIAS}}$ to restore the desired ramp slope. In addition,
distributed interconnect resistance can introduce position-dependent
timing differences between comparator branches. The physical loading
and routing-induced effects of the shared ramp are quantified using
Calibre xACT parasitic extraction in Section~III-D.

\subsection{Variability Analysis}
\begin{figure}[!t]
    \centering
    \includegraphics[width=0.48\textwidth]{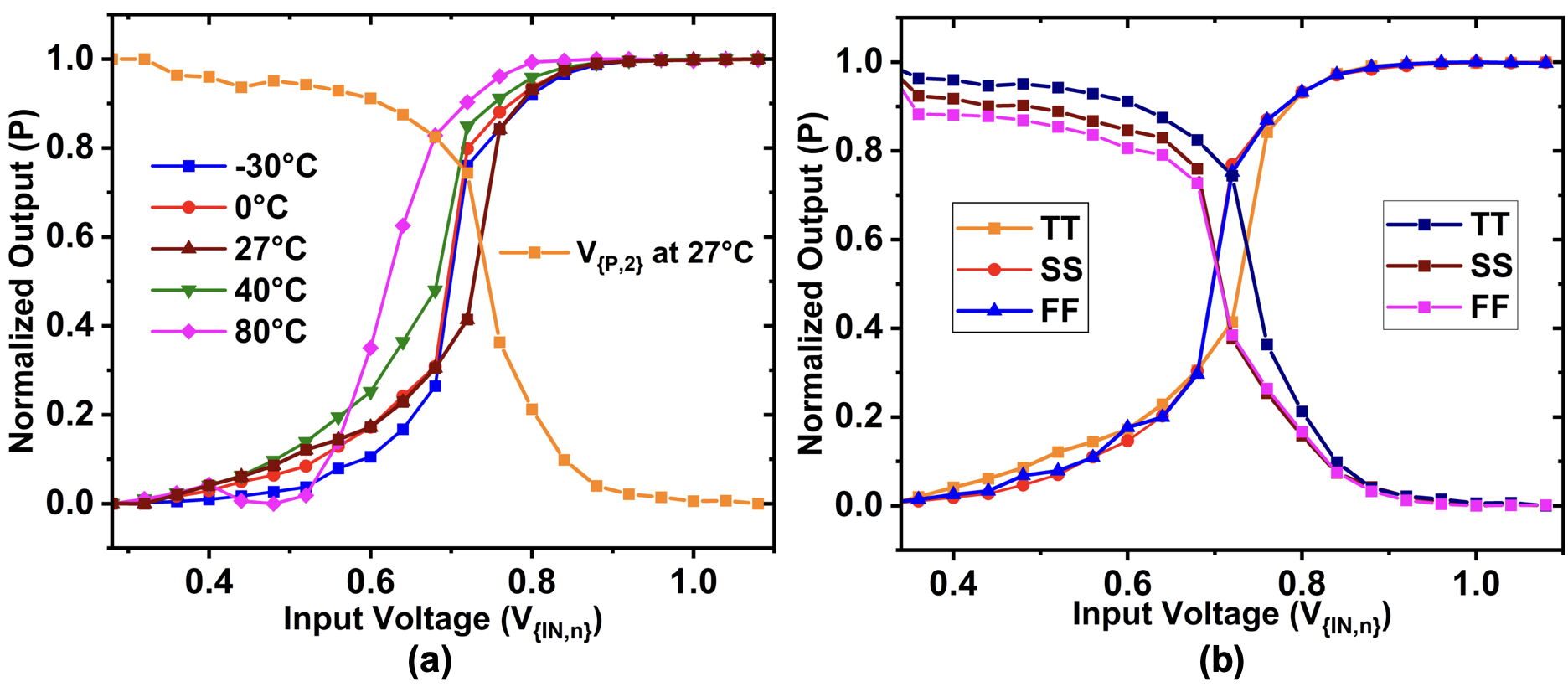}
    \caption{Simulated softmax characteristics: (a) temperature, (b) process curve.}
    \label{fig:tempprocess}
\end{figure}

Cadence Monte Carlo simulations were performed with the full
GF22-FDSOI PDK statistical models enabled, including transistor and modeled passive-component variations. The memristor is excluded from the foundry statistical variation because it is implemented using an external Verilog-A model rather than a native GF22-FDSOI PDK device.
Fig.~\ref{fig:tempprocess}(a)--(b) presents the softmax
transfer characteristics across temperature and process corners for both the swept-channel output ($V_{P,1}$) and the common reference-channel output ($V_{P,n}$). Over temperature,
Fig.~\ref{fig:tempprocess}(a) shows the extracted softmax gain
$\gamma$ varying from 11.36~V$^{-1}$ at $27^\circ$C (TT) to
19.2~V$^{-1}$ at $80^\circ$C, over the temperature range from
$-30^\circ$C to $80^\circ$C.
Across process corners, Fig.~\ref{fig:tempprocess}(b) shows  the corresponding $V_{P,n}$ response exhibits a
consistently tighter agreement with the theoretical characteristic and preserves the complementary softmax behavior, intersecting $V_{P,1}$near the equal-input condition for all corners. 

\begin{figure}[!t]
    \centering
    \includegraphics[width=0.4\textwidth]{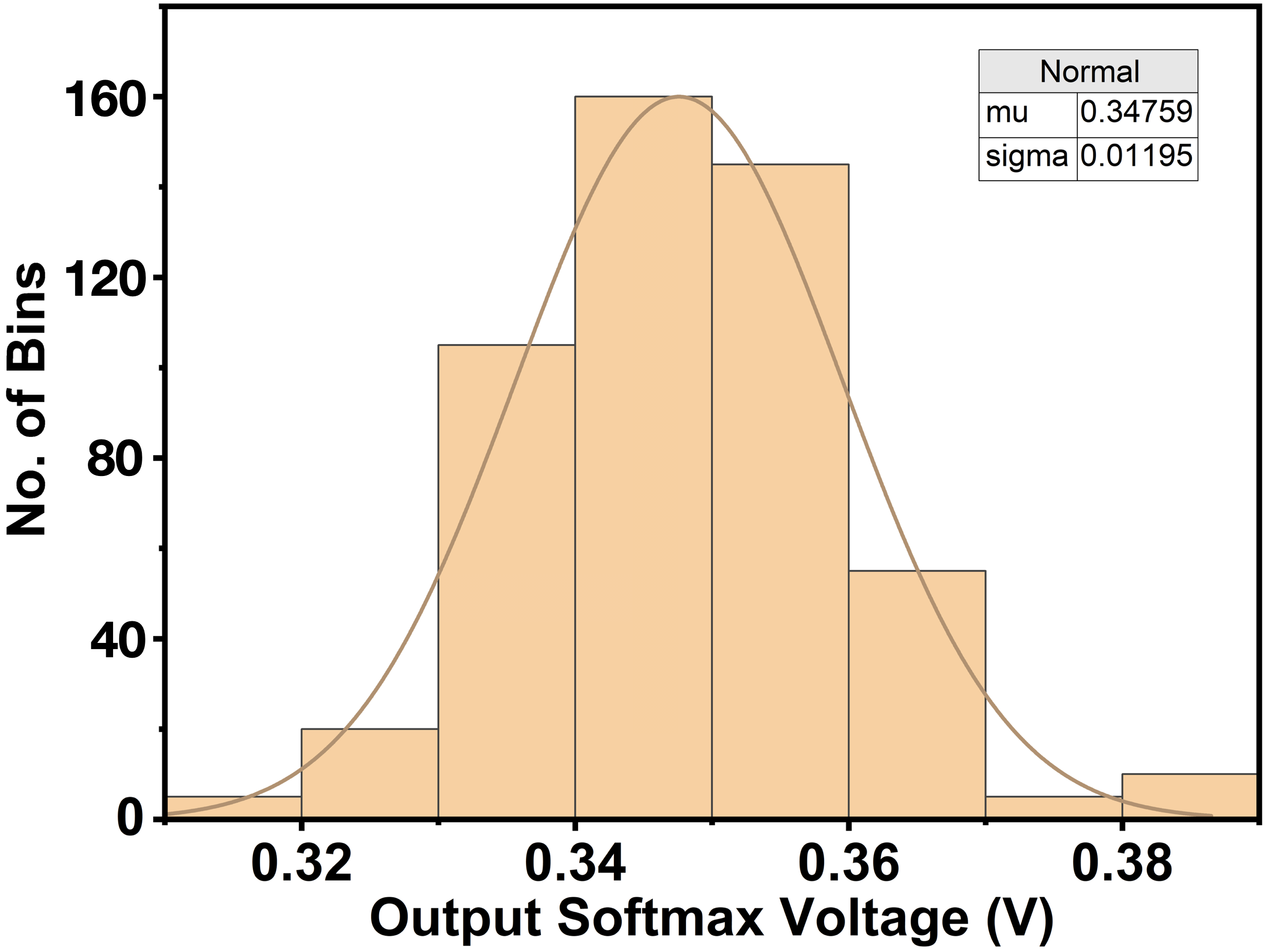}
    \caption{Monte Carlo distribution of the softmax output voltage.}
    \label{fig:montecarlo}
\end{figure}

Fig.~\ref{fig:montecarlo} shows the Monte Carlo distribution of the softmax output voltage. The distribution is approximately Gaussian, with $\mu=0.348$~V and $\sigma=12$~mV, corresponding to a relative spread of approximately 3.4\%. This indicates that the softmax output remains reasonably robust to device-level mismatch variation. As discussed in Section~III-B, the most critical nonidealities arise at
the small-capacitance nodes, where element-to-element capacitance
variation can perturb the sampled voltage and consequently the softmax
output. The sensitivity to such capacitance mismatch was explicitly
evaluated through the array-level $C_C$ and $C_P$ mismatch analysis.
Comparator offset represents another source of error however, it can
be mitigated using standard offset-calibration techniques without
modifying the proposed softmax operation.  Collectively, the monte carlo, temperature, and process-corner results demonstrate that
the proposed circuit maintains the expected softmax behavior across the considered variability conditions.

\subsection{Layout and Post-Layout Extraction Results}

\begin{figure}[!t]
    \centering
    \includegraphics[width=0.48\textwidth]{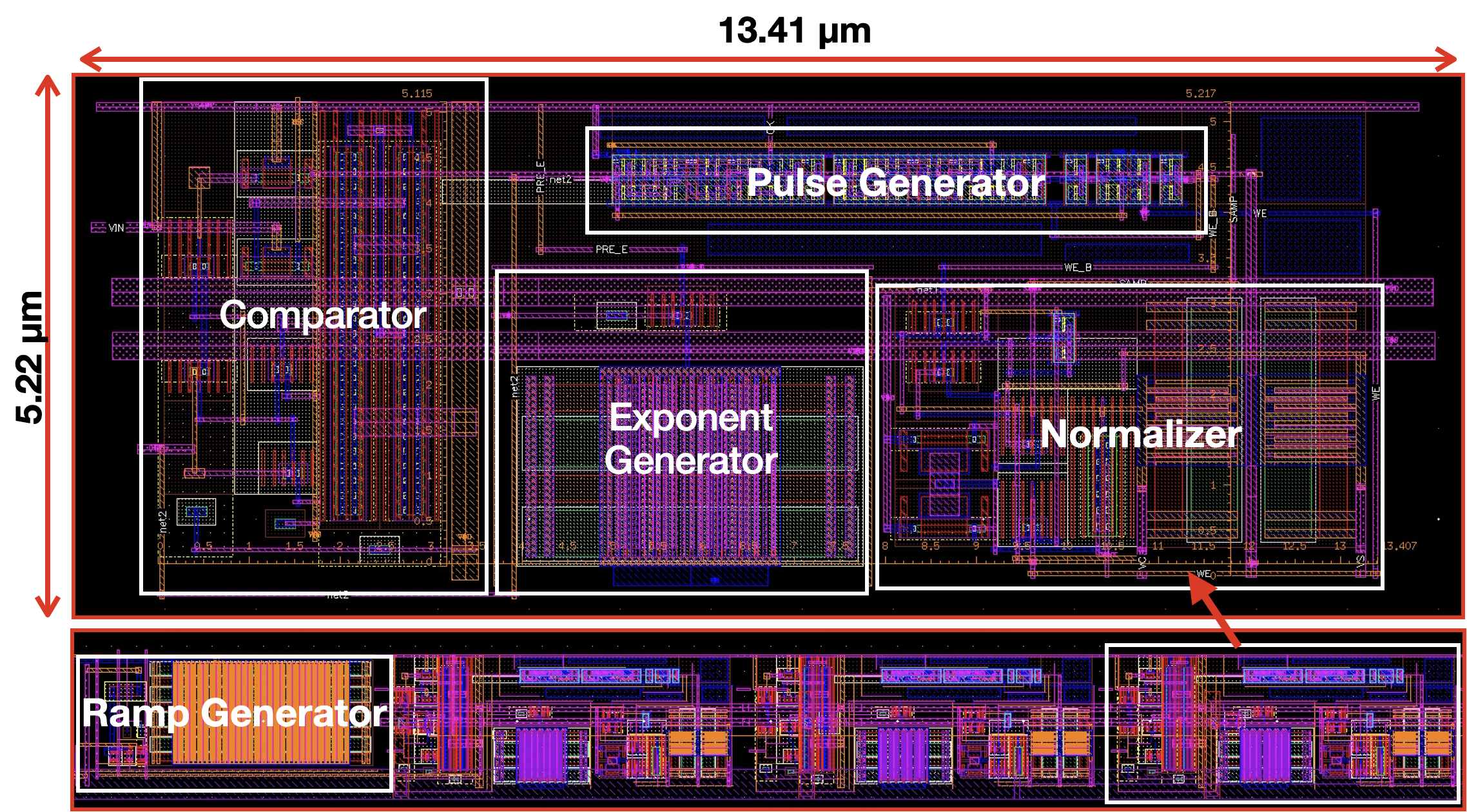}
    \caption{Layout of the proposed design}
    \label{fig:layout}
\end{figure}

To evaluate the physical implementation of the proposed softmax
architecture, a layout of the complete softmax element was implemented
in the GlobalFoundries 22-nm FDSOI technology. The layout of one
softmax element occupies an area of
$13.41~\mu\mathrm{m}\times5.22~\mu\mathrm{m}$, corresponding to
approximately $70.2~\mu\mathrm{m}^{2}$. The complete 128-element
softmax architecture occupies approximately
$1811.5~\mu\mathrm{m}\times5.22~\mu\mathrm{m}$, including the
replicated K elements and shared peripheral circuitry.
Fig.~\ref{fig:layout} shows the layout of the proposed softmax element
and the corresponding 128-element floorplan.

To quantify the physical loading of the globally distributed
$V_{\mathrm{RAMP}}$ network in the 128-channel softmax architecture,
the shared ramp interconnect was routed on C5 using a $1.2~\mu$m wire
width and extracted using Calibre xACT. The extracted network exhibits
approx. $210$~fF of total parasitic capacitance, including ground
and coupling components, together with an interconnect resistance of
approximately $120~\Omega$. The extracted parasitic network captures
the combined loading introduced by the global interconnect, via
connections, and the connected comparator input branches. The complete
distributed RC network is retained in the post-layout simulation such
that both capacitive loading and routing resistance are included in the
circuit-level evaluation.

The shared $V_{\mathrm{RAMP}}$ interconnect was designed using a parasitic-aware approach. Since the long global ramp line provides significant intrinsic capacitance, the explicit ramp capacitor was reduced from the schematic value of $200$~fF to $50$~fF in layout. The extracted interconnect capacitance therefore contributes directly
to the effective ramp capacitance, reducing explicit capacitor area. Any residual shift in ramp slope is compensated through $V_{\mathrm{BIAS}}$.

Since the effective ramp slope is approximately
\begin{equation}
S_{R,\mathrm{eff}}
\approx
\frac{I_R(V_{\mathrm{BIAS}})}
{C_R+C_{\mathrm{RAMP,par}}},
\end{equation}
any residual shift in the extracted ramp slope can be compensated by
adjusting $V_{\mathrm{BIAS}}$. This provides post-layout control of the
effective softmax temperature without modifying the physical ramp
network.

\begin{figure}[!t]
    \centering
    \includegraphics[width=0.48\textwidth]{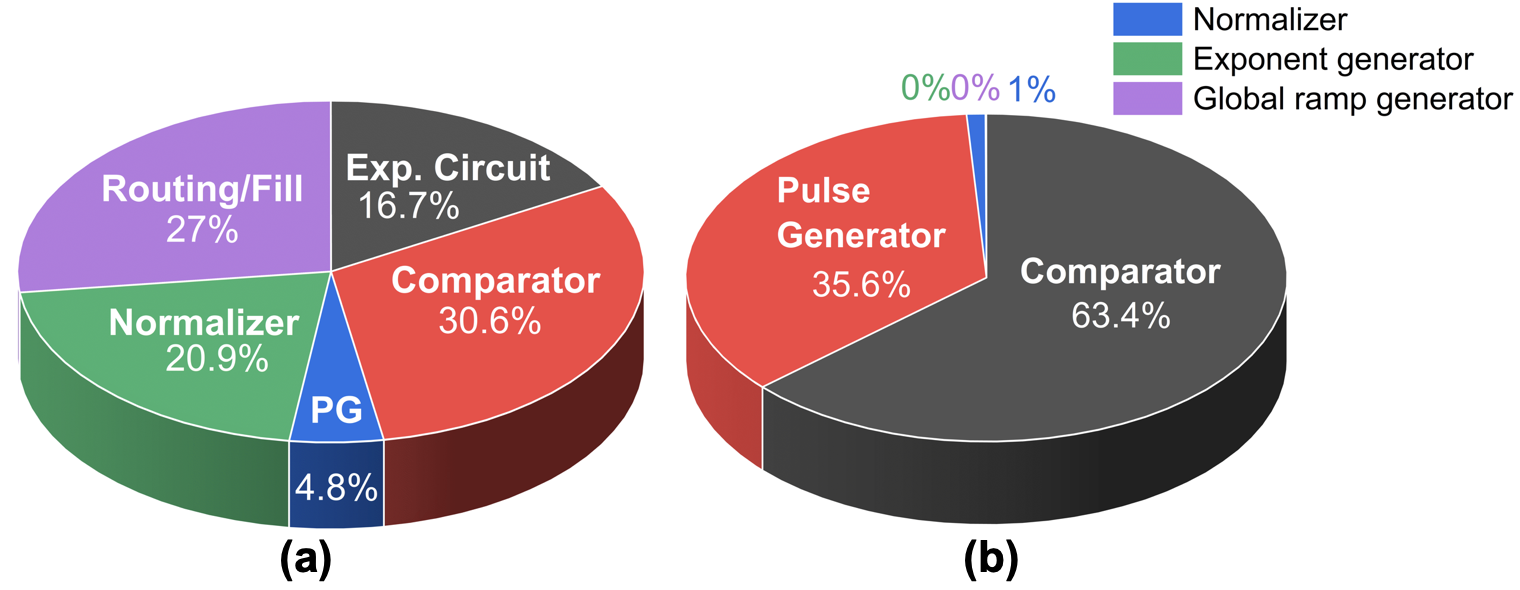}
    \caption{Area and power breakdown of the proposed 128-element softmax architecture: (a) per-element layout area and (b) full-array average power.}
    \label{fig:pie}
\end{figure}

The area contribution of the major circuit blocks is summarized in
Table~\ref{tab:area_breakdown}. The comparator, pulse-generation
circuit, exponential generator, and normalization circuitry constitute
the primary replicated area components, whereas the ramp generator is
implemented once as shared peripheral circuitry.
Fig.~\ref{fig:pie}(a) shows the relative area distribution of the major
per-element circuit blocks. The comparator and normalization circuits
occupy the largest fractions of the softmax-element area, followed by
the exponential generator and pulse generator.
Fig.~\ref{fig:pie}(b) and Table~\ref{tab:power_breakdown} summarize the
power consumption of the complete 128-element softmax architecture.
The comparator and pulse-generation circuits dominate the power
consumption, while the normalizer, exponential generator, and shared
ramp contribute only a small fraction of the total. The complete
architecture consumes approximately $13.44$~mW at
$V_{\mathrm{DD}}=1.1$~V.

\begin{table}[!t]
\centering
\caption{Layout Area Breakdown of the Proposed Softmax Circuit}
\label{tab:area_breakdown}
\renewcommand{\arraystretch}{1.15}
\begin{tabular}{lccc}
\hline
\textbf{Circuit Block} &
\textbf{Length ($\mu$m)} &
\textbf{Width ($\mu$m)} &
\textbf{Area ($\mu$m$^2$)} \\
\hline
Ramp generator          & 12.1 & 4.2 & 50.82 \\
Exponential generator   & 3.80 & 3.08 & 11.70 \\
Comparator              & 4.2 & 5.12 & 21.50 \\
Pulse generator         & 6.28 & 0.54 & 3.39 \\
Normalization circuit   & 5.22 & 2.81 & 14.67 \\
Routing    & -- & -- & 18.94 \\

\hline
\end{tabular}
\end{table}

\begin{table}[!t]
\centering
\caption{Power Breakdown of the 128-Element Softmax Circuit}
\label{tab:power_breakdown}
\renewcommand{\arraystretch}{1.15}
\begin{tabular}{lccc}
\hline
\textbf{Circuit Block} &
\textbf{Count} &
\textbf{Power ($\mu$W)} &
\textbf{Total Power (mW)} \\
\hline
Comparator           & 128 & 66.45  & 8.506  \\
Pulse generator       & 128 & 37.32  & 4.777  \\
Normalizer            & 128 & 1.061  & 0.1358 \\
Exponential generator & 128 & 0.159  & 0.0204 \\
Global ramp generator & 1   & 0.445  & 0.00045 \\
\hline
\textbf{Total} & -- & 105.43 & \textbf{13.44} \\
\hline
\end{tabular}
\end{table}

% \begin{figure}[!t]
%     \centering
%     \includegraphics[width=0.48\textwidth]{Picture/corner_mem.png}
%     \caption{(a) Softmax transfer across process corners (b) MemTorch nanoGPT validation loss, proposed vs.\ ideal softmax.}
%     \label{fig:pvt}
% \end{figure}

\begin{figure}[!t]
    \centering
    \includegraphics[width=0.42\textwidth]{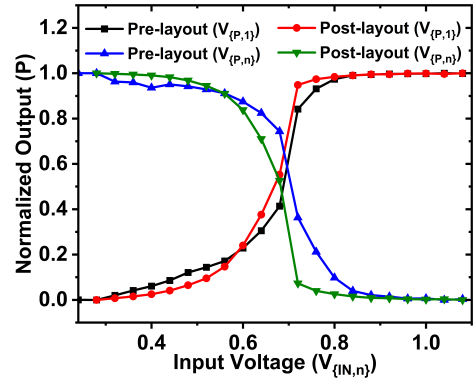}
    \caption{ Pre vs post-layout analysis of the proposed softmax.}
    \label{fig:postlayout}
\end{figure}

Post-layout simulations were performed using the complete
parasitic-extracted network to evaluate whether the physical implementation preserves the nominal softmax characteristic. The extracted results are compared with the schematic-level transfer characteristic previously shown in Fig.~\ref{fig:softmax_transient}(a), where $V_{\mathrm{IN},1}$ is swept while the remaining 127 inputs are held at $0.7$~V. In the schematic simulation, $V_{P,1}$ increases with $V_{\mathrm{IN},1}$, whereas $V_{P,n}$ decreases because all branches share the common normalization denominator, with the two responses intersecting near the equal-input condition $V_{\mathrm{IN},1}=0.7$~V. The same characterization presented in Fig.~\ref{fig:postlayout} using
the extracted circuit to directly quantify the impact of routing and device parasitics.

\begin{figure}[!t]
    \centering
    \includegraphics[width=0.48\textwidth]{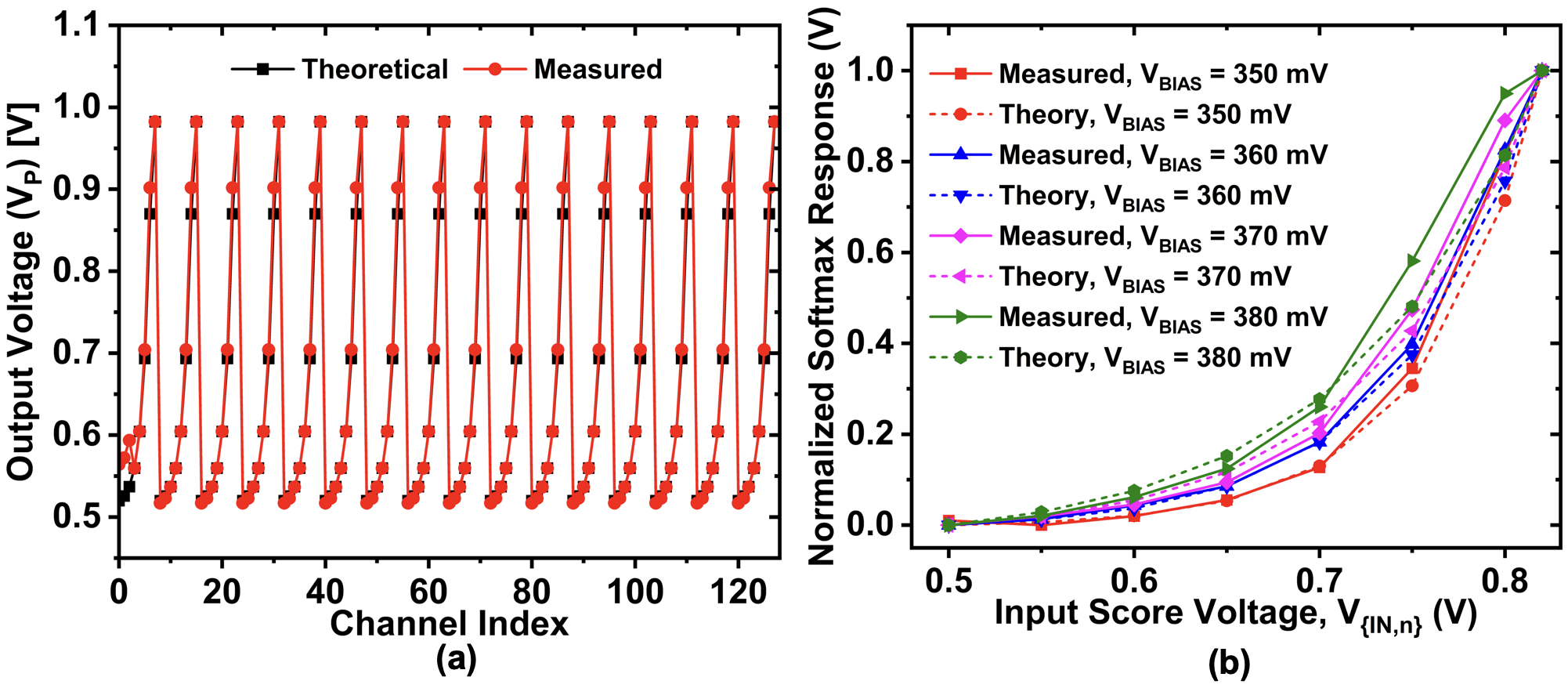}
    \caption{Post-layout extracted programmable softmax response under
    ramp-bias control (a) vector-level validation (b) responses for
    $V_{\mathrm{BIAS}}=350$, 360, 370, and 380~mV are compared with
    theoretical softmax characteristics calculated using the
    corresponding extracted ramp slopes and
    $\gamma_{\mathrm{th}}=2/(S_R\tau_E)$.}
    \label{fig:pex_programmability}
\end{figure}

To further verify that the softmax temperature remains programmable
after parasitic extraction, $V_{\mathrm{BIAS}}$ was varied from
350 to 380~mV using the complete extracted circuit. Increasing
$V_{\mathrm{BIAS}}$ increases the ramp current and reduces the ramp
fall time, resulting in a larger effective ramp slope $S_R$.
According to $\gamma_{\mathrm{th}}=2/(S_R\tau_E)$, the corresponding
softmax gain therefore decreases as the ramp becomes steeper.
Fig.~\ref{fig:pex_programmability}(b) compares the PEX-measured
responses with the theoretical softmax characteristics calculated
using the extracted ramp slopes. The gradual change
in the response confirms that the softmax temperature remains
programmable in the presence of extracted interconnect and device
parasitics. To further verify vector-level operation, the extracted
circuit was evaluated using the complete 128-element interleaved input
vector containing eight voltage levels of $0.50$, $0.55$, $0.60$,
$0.65$, $0.70$, $0.75$, $0.80$, and $0.82$~V.
Fig.~\ref{fig:pex_programmability}(a) shows a representative result at
$V_{\mathrm{BIAS}}=360$~mV, where the PEX-measured output closely
follows the theoretical softmax response obtained using
$\gamma_{\mathrm{th}}=13.72$~V$^{-1}$, yielding an RMSE of
14.12~mV across the 128 channels. The extracted ramp parameters and
corresponding softmax accuracy for the investigated bias conditions are
summarized in Table~\ref{tab:pex_programmability}.

\begin{table}[t]
\centering
\caption{Post-Layout Extracted Softmax Programmability Under Ramp-Bias Control}
\label{tab:pex_programmability}
\begin{tabular}{ccccc}
\hline
$V_{\mathrm{BIAS}}$
& $t_{\mathrm{fall}}$
& $S_R$
& $\gamma_{\mathrm{th}}$
& RMSE \\
(mV) & (ns) & (mV/ns) & (V$^{-1}$) & (mV) \\
\hline
350 & 300 & 3.66 & 14.55 & 18.61 \\
360 & 283 & 3.89 & 13.72 & 14.12 \\
370 & 241 & 4.56 & 11.68 & 21.25 \\
380 & 202 & 5.45 &  9.79 & 29.36 \\
\hline
\end{tabular}
\end{table}

\section{Application and Comparison}

\subsection{Transformer-Level Evaluation Based on MemTorch}

The application-level impact of the proposed softmax architecture was investigated using a hardware-aware Transformer framework implemented with MemTorch \cite{lammie2022memtorch}. A trained nanoGPT model containing approximately 1.8 million parameters was considered, with its linear operations
mapped to memristive crossbar equivalents. The circuit-derived softmax model was then incorporated into the attention path following the $QK^{T}/\sqrt{d_k}$ operation. The ramp-based exponential generation,
saturation-mode normalization, and corresponding inverse-temperature coefficient $\gamma_{\mathrm{th}}$ follow the circuit formulation derived in Section~II.

To interface the dimensionless Transformer attention scores with the analog input range of the proposed circuit, each attention row is first referenced to its maximum score and mapped to the corresponding circuit input voltage according to
\begin{equation}
V_{\mathrm{IN},i}
=
V_H
+
G_{\mathrm{SV}}
\left(S_i-S_{\max}\right),
\label{eq:memtorch_score_mapping}
\end{equation}

where $S_i$ is the $i$th attention score, $S_{\max}$ is the maximum score within the corresponding attention row, $V_H=1.1$~V is the upper limit of the circuit input range, and $G_{\mathrm{SV}}$ denotes the score-to-voltage conversion factor. The mapped voltage is restricted to
the experimentally evaluated circuit range of 0.30--1.10~V. For the nominal hardware-aware evaluation, $G_{\mathrm{SV}}=0.0844$~V/score was used together with the nominal circuit-derived coefficient $\gamma_{\mathrm{th}}=11.85$~V$^{-1}$. The remaining softmax operation follows the circuit behavior described in Section~II. The principal hardware-aware parameters used in the MemTorch evaluation are summarized in Table~\ref{tab:memtorch_softmax}.

\begin{table}[!t]
\centering
\caption{Parameters Used in the MemTorch Hardware-Aware Evaluation}
\label{tab:memtorch_softmax}
\footnotesize
\begin{tabular}{l c}
\hline
\textbf{Parameter} & \textbf{Value} \\
\hline
Softmax input range & 0.30--1.10 V \\
Nominal $\gamma_{\mathrm{th}}$ & 11.85 V$^{-1}$ \\
Score-to-voltage factor $G_{\mathrm{SV}}$ & 0.0844 V/score \\
$V_{\mathrm{DD}}$ & 1.1 V \\
Channel-length modulation $\lambda$ & 0 \\
Output full scale $V_{\mathrm{FS}}$ & 1.0 V \\
Normalizer iterations & 8 \\
Normalizer current model & Square-law \\
\hline
\end{tabular}
\end{table}

\begin{figure}[!t]
    \centering
    \includegraphics[width=0.48\textwidth]{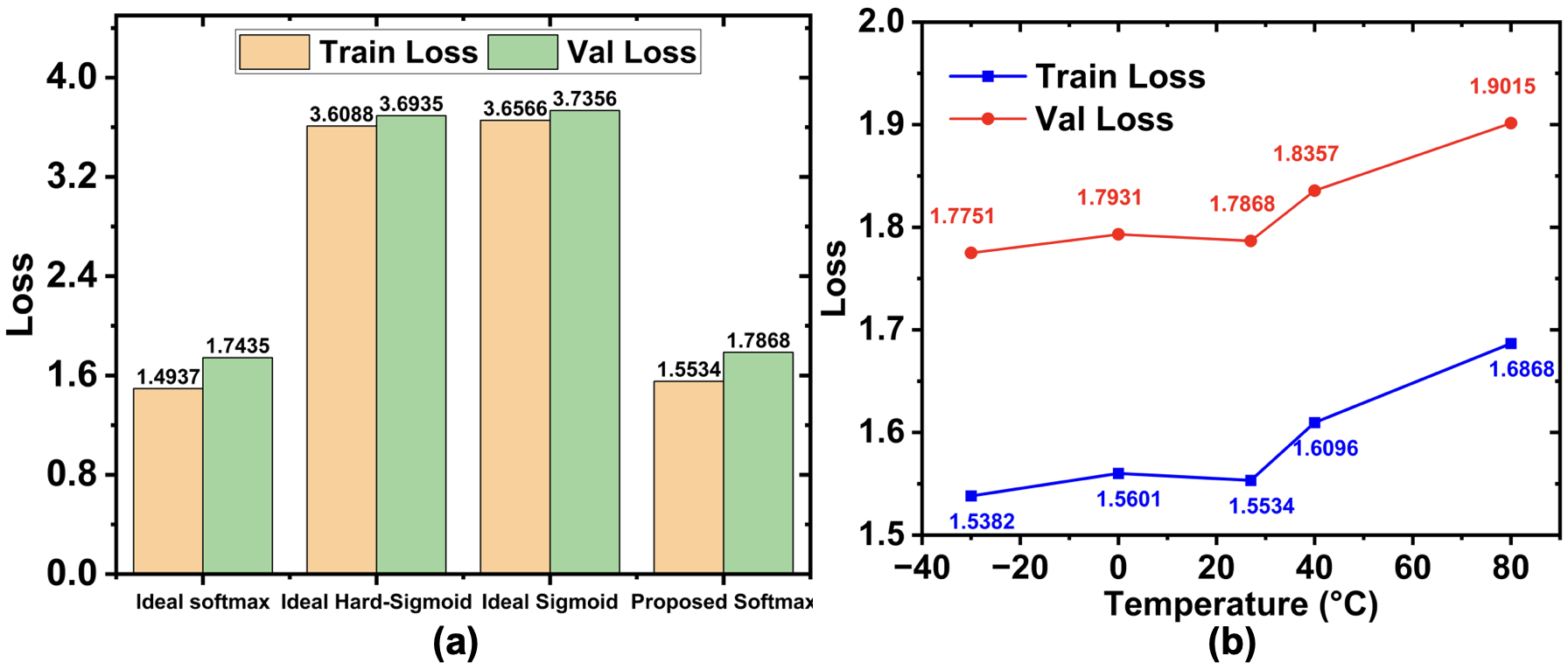}
    \caption{Hardware-aware Transformer evaluation. (a) Loss comparison of ideal and hardware softmax alternatives. (b) Temperature-dependent loss of the proposed softmax for different $\gamma_{\mathrm{th}}$.}
    \label{fig:memtorch_results}
\end{figure}

Fig.~\ref{fig:memtorch_results}(a) compares the Transformer training and validation losses obtained using the ideal softmax, sigmoid, hard-sigmoid, and the proposed hardware-aware softmax. The ideal softmax achieves training and validation losses of 1.4937 and 1.7435,
respectively, whereas the proposed softmax results in corresponding losses of 1.5534 and 1.7868. Thus, the proposed circuit introduces only approximately a 2.48\% increase in validation loss relative to the ideal softmax. In contrast, sigmoid and hard-sigmoid replacements
increase the validation loss to 3.7356 and 3.6935, respectively, indicating substantially larger degradation of the attention behavior. These results show that the proposed hardware-aware softmax preserves the Transformer-level response much more closely than the simpler nonlinear replacements. 

The impact of circuit-level temperature variation was further evaluated by propagating the extracted inverse-temperature coefficients into the hardware-aware Transformer model. As shown in Fig.~\ref{fig:memtorch_results}(b), the circuit-derived $\gamma_{\mathrm{th}}$ values are 12.0, 13.1, 11.36, 15.6, and 19.2~V$^{-1}$ at $-30$, 0, 27, 40, and $80^\circ$C, respectively. The corresponding validation losses are 1.7751, 1.7931, 1.7868, 1.8357, and 1.9015. The nominal $27^\circ$C condition therefore remains close to the proposed-softmax result, while the largest degradation
occurs at $80^\circ$C due to the increased circuit-derived
$\gamma_{\mathrm{th}}$. Nevertheless, the proposed hardware softmax maintains considerably lower loss than the sigmoid and hard-sigmoid alternatives across the evaluated temperature range.

\newcommand{\centercell}[1]{%
\begin{tabular}[c]{@{}c@{}}
\rule{0pt}{0.5ex}#1\rule[-1.0ex]{0pt}{0.5ex}
\end{tabular}}

\begin{table*}[t]
\centering
\scriptsize
\setlength{\tabcolsep}{2.2pt}
\renewcommand{\arraystretch}{1.25}

\caption{Comparison With State-of-the-Art Softmax Hardware Implementations}
\label{tab:comparison}

\begin{tabular}{|c|c|c|c|c|c|c|c|c|c|c|}
\hline

\textbf{Metric}
& \textbf{\centercell{Vatalaro\\\cite{vatalaro2021low}}}
& \textbf{\centercell{Jiang\\\cite{jiang2026self}}}
& \textbf{\centercell{Kouretas\\\cite{kouretas2019simplified}}}
& \textbf{\centercell{Spagnolo\\\cite{spagnolo2021aggressive}}}
& \textbf{\centercell{Base-2\\\cite{zhang2022base}}}
& \textbf{\centercell{SoftAct\\\cite{fu2024softact}}}
& \textbf{\centercell{Kim\\\cite{kim2024hardware}}}
& \textbf{\centercell{Wang\\\cite{wang2018high}}}
& \textbf{\centercell{Hyft\\\cite{xia2024hyft}}}
& \textbf{This Work} \\
\hline

\textbf{Technology}
& 180 nm
& 28 nm
& 90 nm
& \centercell{28 nm\\FDSOI}
& 28 nm
& 28 nm
& 28 nm
& 28 nm
& \centercell{FPGA\\Zynq}
& \textbf{\centercell{22 nm\\FDSOI}} \\
\hline

\textbf{Domain}
& Analog
& Analog
& Digital
& Digital
& Digital
& Digital
& Digital
& Digital
& Digital
& \textbf{\centercell{Analog\\time-domain}} \\
\hline

\textbf{\centercell{Exponential\\mechanism}}
& Subthresh.
& Subthresh.
& LUT
& \centercell{Shift-add\\approx.}
& SM-LUT
& \centercell{PLF +\\penalty}
& \centercell{Bit-wise\\LUT}
& Linear fit
& \centercell{Hybrid\\fixed/float}
& \textbf{\centercell{Ramp +\\RC decay}} \\
\hline

\textbf{Precision}
& Analog
& \centercell{Analog\\4-b MAC\\8-b ADC}
& \centercell{10-b\\fixed}
& \centercell{16-b\\fixed}
& \centercell{16-b\\fixed}
& \centercell{16-b\\fixed}
& \centercell{8-b\\fixed}
& \centercell{8-/16-b\\fixed}
& FP16/FP32
& \textbf{Analog} \\
\hline

\textbf{Parallelism}
& 10
& 4
& 10
& 10
& 10
& 1
& 8
& 16/8
& 128
& \textbf{128} \\
\hline

\textbf{\centercell{Computation\\mode}}
& Parallel
& Parallel
& \centercell{Parallel\\comb.}
& \centercell{Parallel}
& Parallel
& Serial
& Serial
& Serial
& Serial
& \textbf{Parallel} \\
\hline

\textbf{Supply (V)}
& 0.5
& 0.9--1.8
& 1.0
& 1.0
& N.R.
& 0.9
& 1.0
& N.R.
& N.R.
& \textbf{1.1} \\
\hline

\textbf{\centercell{Reported\\vector size (N)}}
& 10
& 4
& 10
& 10
& 10
& 1
& 8
& 16
& 128
& \textbf{128} \\
\hline

\textbf{\centercell{Total area\\($\mu$m$^2$)}}
& 2570
& 1361
& 16,891
& 3595
& 5676
& 2872
& 7100
& 15,000
& \centercell{41,116\\LUT+FF}
& \textbf{9453.42} \\
\hline

\textbf{\centercell{Area/element\\($\mu$m$^2$)}}
& 257
& 340
& 1689
& 360
& 568
& 2872
& 888
& 938
& \centercell{321\\LUT+FF}
& \textbf{70.2} \\
\hline

\textbf{\centercell{Total power\\(mW)}}
& 0.003
& N.R.
& 1.612
& 24
& 13.12
& 3.46
& 22.82
& 51.60
& 4920
& \textbf{13.44} \\
\hline

\textbf{\centercell{Power/element\\($\mu$W)}}
& 0.3
& N.R.
& 161.2
& 2400
& 1310
& 3460
& 2850
& 3230
& 38,400
& \textbf{105} \\
\hline

\textbf{\centercell{Energy/element\\(pJ)}}
& 1.02
& N.R.
& 0.47
& 0.96$^{\dag}$
& 0.44
& 1.87$^{*}$
& 0.91
& 1.15$^{*}$
& 1073
& \textbf{25.5} \\
\hline

\textbf{\centercell{Latency\\(ns)}}
& \centercell{3410\\single-shot}
& \centercell{25}
& 2.93
& 1.2
& N.R.
& 141$^{*}$
& $\sim$15
& 5.7$^{*}$
& 27.9
& \textbf{\centercell{242.97\\single-shot}} \\
\hline

\textbf{\centercell{Programmable\\temperature}}
& \centercell{Yes\\$V_{DD}$}
& No
& No
& No
& No
& No
& No
& No
& No
& \textbf{\centercell{Yes\\ramp + RC}} \\
\hline

\textbf{\centercell{System-level\\validation}}
& None
& \centercell{MobileViT-\\XXS}
& \centercell{LeNet/VGG/\\ResNet}
& \centercell{LeNet/VGG/\\ResNet/CIFAR}
& \centercell{ResNet/\\CIFAR}
& \centercell{MobileViT-\\XXS}
& \centercell{Random\\vectors}
& LeNet-5
& \centercell{BERT/GLUE/\\SQuAD}
& \textbf{\centercell{Transformer\\MemTorch}} \\
\hline

\end{tabular}

\vspace{3pt}
\raggedright

$^{\dag}$ Directly reported by the source, not derived.
$^{*}$ Computed from reported power and throughput; excludes pipeline
fill and one-time stages such as logarithm or reciprocal evaluation.
N.R.: not reported.

\end{table*}

\subsection{Comparison with Prior Softmax Hardware}

The proposed softmax array was evaluated for the 128 column outputs of a
$128\times128$ vector--matrix multiplication (VMM) array. Each VMM
output column is connected to one softmax branch therefore, the 128
branches collectively perform one 128-element softmax evaluation. The
complete array, including precharge, exponential-weight generation,
sampling, and normalization, operates in parallel and completes a
128-element softmax operation in 242.97~ns. The total power consumption of
the 128-branch array is 13.44~mW, corresponding to approximately
105~$\mu$W per element. Using the single-shot evaluation latency, the
total energy per 128-element softmax evaluation is approximately
3.27~nJ, or 25.5~pJ per output element.

Table~\ref{tab:comparison} compares the proposed architecture with
state-of-the-art analog and digital softmax hardware implementations.
Because the reported vector sizes and degrees of parallelism differ
considerably among prior works, both total and per-element area, power,
and energy are included to provide a more direct comparison. The
proposed architecture supports 128-way parallel operation, whereas most
of the custom ASIC implementations \cite{vatalaro2021low,jiang2026self,kouretas2019simplified,spagnolo2021aggressive,zhang2022base,fu2024softact,kim2024hardware,wang2018high} in Table~\ref{tab:comparison}
evaluate vectors of 1--16 elements. Despite implementing 128 parallel
softmax branches, the proposed design occupies 9453.42~$\mu$m$^2$,
corresponding to 70~$\mu$m$^2$ per element, which is the smallest
per-element area among the compared implementations \cite{vatalaro2021low,jiang2026self,kouretas2019simplified,spagnolo2021aggressive,zhang2022base,fu2024softact,kim2024hardware,wang2018high,xia2024hyft}. The architecture
also achieves a per-element power of approximately 105~$\mu$W while
performing the complete 128-element operation in a single 242.97-ns
evaluation.

Unlike digital implementations that approximate the exponential
function using LUTs, shift--add operations, piecewise-linear functions,
or other arithmetic approximations \cite{kouretas2019simplified,zhang2022base,kim2024hardware}, the proposed circuit directly
generates the exponential response in the analog time domain using a
shared ramp and RC decay. Earlier analog softmax circuits
typically exploit MOSFET subthreshold behavior, for which the
exponential characteristic is confined to a relatively narrow operating
range and is sensitive to device and temperature variations
~\cite{vatalaro2021low,jiang2026self}. In contrast, the proposed
exponential generator is governed by the ramp swing and RC time
constant, allowing operation over the demonstrated 0.30--1.10~V input
range. Moreover, the effective softmax temperature can be programmed
through the ramp slope and RC time constant, providing an additional
degree of control that is unavailable in most of the compared
implementations. Finally, the proposed architecture is evaluated at the
system level using a Transformer with MemTorch, demonstrating its
applicability beyond isolated circuit-level softmax evaluation.
% \ag{1. Charge injection experiment: For a single branch, plot VExp and VE, zoomed in around switching point and check the charge injection and its effect on VE, do it for all ranges of VEXP value.}

% \ag{2. Continuing from 1, Plot VE after say 100 ns.}

% \ag{3. Show the characterization of the comparator offset.}

\subsection{Future Works and Scalability}

Future work will investigate the scalability of the proposed architecture
to softmax dimensions beyond 128 elements. As the number of branches
increases, the parasitic loading of the shared $V_{\mathrm{RAMP}}$
interconnect is expected to increase, which can alter the ramp slope and
the corresponding $\gamma_{\mathrm{th}}$. This can be addressed through
hierarchical partitioning of the softmax array, where smaller groups of
branches employ replicated ramp-driving circuits while operating in
parallel. In addition, the proposed softmax architecture can be
integrated with physical crossbar-based VMM arrays to enable complete
hardware-level evaluation of Transformer operations. Such an
implementation would allow the effects of device and circuit
nonidealities, data-conversion overhead, and interface circuitry to be
evaluated jointly with application-level Transformer accuracy. The memristive element used in the exponential-decay path is modeled
using an external Verilog-A model because a memristor device is
not available in the employed GF22-FDSOI PDK. In a compatible resistive-memory technology, the
memristor can be integrated in the back-end-of-line (BEOL) interconnect
stack above the underlying CMOS circuitry. Since such a device can be
vertically integrated without requiring a dedicated front-end-of-line (FEOL) silicon region, its inclusion is expected to cause only a limited increase in the effective implementation footprint, primarily due to routing and
integration constraints. Hence, the overall per-element area is expected
to remain comparable to the CMOS footprint reported in this work.

\section{Conclusion}
This work presents a 128-channel time-domain analog softmax architecture
for CIM output processing using a shared falling ramp, RC-decay-based
exponential generation, and in-circuit normalization. The circuit
supports a wide 0.30--1.10~V input range and provides programmable
softmax temperature through the ramp slope and RC time constant.
Post-layout extracted simulations verify the operation of the complete
128-channel array, which occupies approximately 9453.42~$\mu$m$^2$,
corresponding to 70.2~$\mu$m$^2$ per element. The array performs all
128 softmax outputs in parallel with a latency of 242.97~ns at
13.44~mW, corresponding to approximately 25.5~pJ per output element.
The circuit also demonstrates robustness under capacitance variation,
process, monte carlo, and extracted interconnect parasitics. MemTorch-based
Transformer evaluation further demonstrates its applicability to
CIM-based attention processing. Future work will focus on integration
with physical VMM arrays and complete hardware-level Transformer
evaluation.

\bibliographystyle{IEEEtran}

\bibliography{sample-base}

\end{document}